# Measures and Models of Brain-Heart Interactions

Diego Candia-Rivera*, Luca Faes, Fabrizio de Vico Fallani, Mario Chavez

*Abstract*—The exploration of brain-heart interactions within various paradigms, including affective computing, human-computer interfaces, and sensorimotor evaluation, stands as a significant milestone in biomarker development and neuroscientific research. A range of techniques, spanning from molecular to behavioral approaches, has been proposed to measure these interactions. Different frameworks use signal processing techniques, from the estimation of brain responses to individual heartbeats to higher-order dynamics linking cardiac inputs to changes in brain organization. This review provides an overview to the most notable signal processing strategies currently used for measuring and modeling brain-heart interactions. It discusses their usability and highlights the main challenges that need to be addressed in future methodological developments. Current methodologies have deepened our understanding of the impact of neural disruptions on brain-heart interactions, solidifying it as a biomarker for evaluation of the physiological state of the nervous system and holding immense potential for disease stratification. The vast outlook of these methods becomes apparent specially in neurological and psychiatric disorders. As we tackle new methodological challenges, gaining a more profound understanding of how these interactions operate, we anticipate further insights into the role of peripheral neurons and the environmental input from the rest of the body in shaping brain functioning.

## I. INTRODUCTION

As early as 1938, evidence suggesting a functional brain-heart interaction was reported in a patient with a brain injury, showing distinctive electrocardiography patterns [1]. Since then, numerous clinical cases have provided abundant evidence linking cardiovascular, neurological, and psychiatric disorders to changes in the brain-heart interaction. For example, severe brain damage can lead to sudden cardiac death [2], while cardiac arrhythmias can cause cerebrovascular accidents such as ischemic attacks [3].

The brain and heart communicate with each other to participate in various processes involved in sensing, integration, and regulation of bodily activity [4], [5], namely interoception. This communication is essential for maintaining neural homeostasis and the overall physiological state of the body [6].

The interoceptive mechanisms operating within the brain-heart axis span various components (Fig. 1), from genetic factors, molecular mechanisms, hormonal and neural pathways [7]. Evidence on genetic factors come from the links between genomic loci associated with both cardiac and brain anatomy, but also between cardiovascular issues and genetic risk for psychiatric disorders, such as major depression, schizophrenia, and bipolar disorder, emphasizing the association between brain function and increased cardiovascular risks [8], [9], [10].

Brain-heart interaction can occur through cellular mechanisms involving extracellular vesicles [11]. In the context of stroke, there is evidence indicating that it can elevate the levels of circulating extracellular vesicles [12]. Additionally, it has the potential to increase the permeability of the blood-brain barrier [13], eventually leading to posterior cardiac dysfunctions [14]. Stroke may also cause the downregulation of certain microRNAs [15], which are non-coding RNAs that play important roles in regulating gene expression, and their transportation through extracellular vesicles may likely target and influence heart physiology. Conversely, cardiac damage can trigger protein-specific release that can induce thrombosis [16], but also alter the regulation of gene expression at brain level [17].

The pulsations of the heart during each beat create mechanical and electromagnetic effects in the brain. The mechanical force generated by the heartbeat sends pressure waves through the blood vessels, influencing cerebral blood flow and promoting efficient oxygen and nutrient delivery, but also influencing neural dynamics [18], which is reported to be mediated by mechanosensitive ion channels. The expression of these ion channels occurs in sensory neurons contributing to the baroreflex, a mechanism to regulate blood pressure [19]. Simultaneously, the electrical activity generated by the heart produces electromagnetic fields at brain level [20] that can influence neural oscillations, but also heartbeats can reach cortical and subcortical structures through different neural pathways. These pathways include visceroceptive and spino-thalamocortical pathways [21], which are in part mediated by the autonomic nervous system through its sympathetic and parasympathetic branches [4].

State-of-the-art of noninvasive methodologies for estimating these brain-heart interactions in humans include different approaches. One approach involves measuring transient neural responses to heartbeats, known as heartbeat-evoked responses,

D.C.R. is supported by the European Commission, Horizon MSCA Program (grant n° 101151118).

D.C.R., F.D.V.F. and M.C. are with the Sorbonne Université, Paris Brain Institute (ICM), CNRS UMR7225, INRIA Paris, INSERM U1127, AP-HP

Hôpital Pitié-Salpêtrière, 75013, Paris, France. L. F. is with the Department of Engineering, University of Palermo, 90128 Palermo, Italy. *Correspondence to D.C.R. (e-mail: diego.candia.r@ug.uchile.cl).

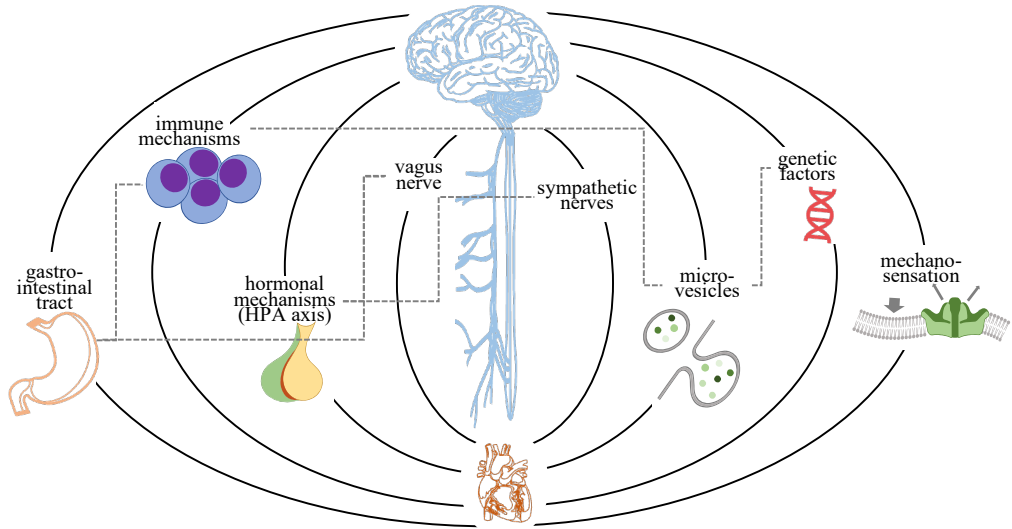

Fig. 1. Pathways of the brain-heart connection. These pathways, which facilitate direct or indirect interactions between the brain and heart, encompass various physiological systems beyond the commonly discussed vagus nerve and sympathetic nerves of the autonomic nervous system. Additional pathways involve hormonal mechanisms within the Hypothalamic-Pituitary-Adrenal axis, and immune mechanisms primarily linked to neuroinflammatory processes initiated by the brain and affecting the heart. The gastrointestinal tract contributes through mechanisms related to the innervation of gut pacemaker cells by parts of the autonomic nervous system, as well as through gut-mediated effects associated with microbiota and gut dysbiosis, which are implicated in conditions such as stroke. Interorgan communication is also facilitated through microvesicles, which contain gene regulation messengers such as microRNAs. Recently, common genetic factors have been identified in brain and heart pathologies, although the mechanisms involved require further elucidation, with some likely associated with genetic regulation through microRNAs. Mechanosensation is another mechanism of brain-heart communication, evidenced by baroreceptor mechanisms and mechanosensitive ion channels that respond to each pulsation.

which relies solely on analyzing brain activity in between heartbeats. Experimental findings demonstrate that heartbeats and the associated cardiac cycle have an impact on perception, information processing, and reaction [22]. It is hypothesized that cardiac inputs to the brain influence the generation of spontaneous cognition, which involves developing a first-person perspective [23]. Therefore, in recent years, there has been a growing call for a paradigm shift in neuroscience research towards an embodied perspective that includes visceral activity.

Other proposals use signal processing techniques to examine correlation, co-occurrences, directional coupling, and higher-order interactions between brain and cardiac autonomic dynamics. Some of these frameworks exploit time-dependent interactions and information theory-based measures [24], [25], [26], [27]. Additionally, further methodologies like Granger Causality, Information Transfer, and Convergent Cross-Mapping have been proposed to assess directionality between the two studied brain and cardiovascular time series [25], [28]. More recent methods focus on analyzing neural systems using generative models of brain and cardiac dynamics. These models leverage prior physiological knowledge and establish connections between changes in brain and heartbeat dynamics. By considering brain and heartbeat oscillations within a framework of mutual influence, these methods offer a potential for causality assessment as well [29], [30]. In addition, certain frameworks focus on measuring complex interactions among various brain regions, occurring simultaneously with autonomic processes. These frameworks explore how cardiac activity may impact brain dynamics and how brain networks evolve in response to physiological fluctuations from other organs.

This review offers a comprehensive look at strategies for measuring and modeling brain-heart interactions. Many of these strategies are versatile and can be extended to study interactions with other bodily systems, including gastric rhythms, respiration, skin conductance, or body temperature. The development and application of these methodologies in different contexts may help to elucidate the physiological underpinnings of the appropriate processing of interoceptive inputs, which plays a crucial role in maintaining a healthy brain. To achieve this, we analyze the practicality of these methods, address current methodological challenges, and outline the most notable clinical translations.

## II. COUPLING BEHAVIOR AND NEURAL ACTIVITY WITH THE CARDIAC CYCLE

The cardiac cycle consists of two phases: systole, the muscle contraction phase; and diastole, the relaxation phase (Fig. 2a). Experimental findings reveal that the phase of the cardiac cycle is associated with perceptual awareness and behavior [22]. Specifically, humans are more likely to detect a stimulus when it is presented during diastole. These findings have been observed in tasks involving visual [31], auditory [32], and somatosensory detection [33], [34]. Conversely, processes such as saccades during visual search [35], [36], visual attention [37], [38], active information sampling [39], active tactile discrimintation [40], reaction time and motor excitability [41], [42], [43], [44], [45] are enhanced during the systole phase. In the study of perceptual awareness and behavior concerning the cardiac cycle phase, synchronizing neural dynamics with the

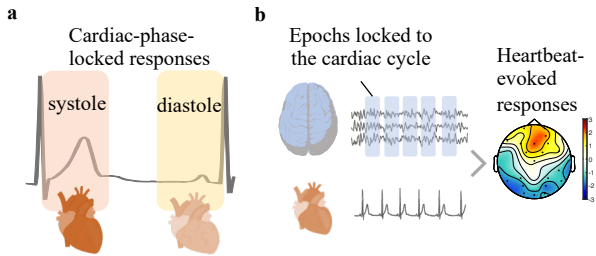

Fig. 2. Measures of brain-heart interaction based on changes in behavioral responses and brain activity with respect to the cardiac cycle. (a) Cardiac phase methods aim at contrasting responses occurring in the systole and diastole phases of the cardiac cycle. (b) Heartbeat-evoked responses aim at providing a signature of the evoked brain responses to individual heartbeats by averaging brain epoch with respect a defined phase of the cardiac cycle.

Central autonomic network

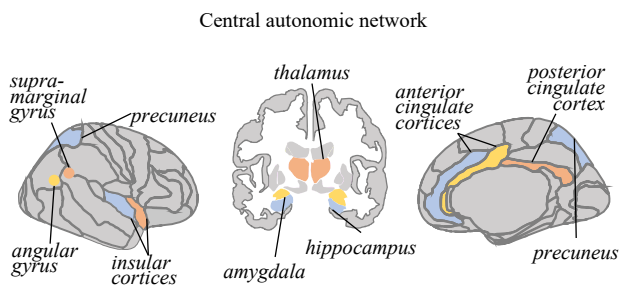

Fig. 3. Central autonomic network components, based on meta-analysis of autonomic correlates [66]: parietal lobe substructures, including the precuneus, angular gyrus, and supramarginal gyrus; anterior and posterior insular cortices; subgenual, pregenual and dorsal anterior cingulate cortices; posterior cingulate cortex; and subcortical structures, including the thalamus, amygdala and hippocampus.

approximate onset of systole and diastole emerges as a compelling approach for analyzing brain-heart interactions [46]. This approach has recently been suggested for extension into fMRI analysis, to timely present stimuli as a function of the cardiac phase [47].

The analysis of brain responses to heartbeats using heartbeat-evoked potentials was initially proposed by Schandry and colleagues in 1986 [48]. Typically, the computation of these potentials involves averaging brain signals that are time-locked to the R- or T-peak of the cardiac cycle [49] (Fig. 2b). However, there is currently no consensus on how to compute them, including aspects like baseline correction, cardiac-field artifact removal, and overall preprocessing [49]. Heartbeat-evoked potentials have been linked to markers of cortical processing of cardiac signals, as they are modulated in various conditions, such as perceptual awareness in healthy state [23], but also in clinical conditions [50], [51], [52], [53], [54], [55]. Though, there is considerable diversity in the specific latencies relative to the cardiac cycle and the scalp locations where these effects are observed [56]. To uncover the intricacies of heartbeat-evoked potentials, further methodological analyses have been proposed to highlight these biomarkers, including information-based techniques [57], time-frequency analysis [58], [59], [60], variability [52], complexity and network properties [61].

The primary limitation of cardiac cycle-based approaches lies in the dynamic variations of sympathetic and parasympathetic autonomic activities influencing the cardiac cycle itself. These variations may intricately connect with brain dynamics in both afferent and efferent manners, leading to a lack of specificity regarding the involved physiological dynamics. Some intracranial studies have identified specific brain regions, such as the anterior cingulate, right insula, prefrontal cortex, and left secondary somatosensory cortex [62], [63], as origins for heartbeat-evoked potentials. However, identifying the cortical and sub-cortical regions involved in heartbeat-evoked potentials using non-invasive techniques remains challenging.

Moreover, limitations also arise from cardiac electric currents associated with ventricular contractions [20], which can induce artifacts in computing heartbeat-evoked potentials. Importantly, it remains to be further elucidated wheter the heartbeat-evoked responses have a direct relationship with the recently uncovered pathways from the mechanical effects of the brain due to changes in blood pressure caused by each heartbeat [22], where recent research in rodent models has shed light on this aspect [18].

Current protocols to study heartbeat-evoked responses typically set a fixed value of latencies and duration for each heartbeat to define a baseline. This approach can, however, be biased by inter-beat variability due to the heart rate variability of the subject during the task. Although some startegies can mitigate the effect of this variability (e.g. discarding short intervals, or adapting the baseline to the events cycles), the conventional procedure for computing and analysing ECG-based cortical responses is rarely questioned.

## III. CO-OCCURRENCES OF AUTONOMIC DYNAMICS IN THE BRAIN

Links between the brain function assessed in specific regions and cardiac rhythmicity have been reported in both neuroimaging and intracranial studies. The associations of autonomic and brain region activities in healthy subjects have been related to direct autonomic control, although the causal relationship is not directly assessed. The estimation of the sympathetic and parasympathetic activities is traditionally done through heart rate variability (HRV) spectral integration at low (LF: 0.04–0.15 Hz) and high frequencies (HF: 0.15–0.4 Hz), respectively [64], although with some variability in the bands' definitions. Because some studies have shown that the estimation of sympathetic activity from HRV can be biased [65], sympathetic markers are also gathered from other physiological activity, such as sympathetic nerve neurogram or electrodermal activity [66].

Neuroimaging evidence increasingly supports the involvement of cortical regions in autonomic dynamics, alongside hypothalamic and brainstem nuclei [67]. Meta-analyses on fMRI studies [66], [68] revealed that the most reported brain regions involved in autonomic correlates are thalamus, hippocampus, amygdala, right anterior insula, left posterior insula, cingulate cortices and a few structures from parietal lobes (Fig. 3). Intracranial electrophysiological recordings have further confirmed the involvement of the anterior and posterior insula, along with limbic system components such as the amygdala, hippocampus, and anterior and mid-cingulate regions [69]. Altogether, those evidences highlight the involvement of numerous high-order regions and the forebrain, but also and several nuclei in the medulla, such as the nucleus of tractus solitarius, nucleus ambiguous, parabrachial Kolliker fuse nucleus [2], [70]; but also in the cerebellum [71], [72]. Further regions have been described in relationship to complex HRV patters [70], including temporal gyrus, planum temporale, frontal orbital cortex, opercular cortex, paracingulate gyri, cingulate gyri, temporal fusiform, superior and middle frontal gyri, lateral occipital cortex, angular gyrus, precuneus cortex, frontal pole, intra-calcarine, supra-calcarine cortices; although, lacking of specificity with respect to their sympathetic or parasympathetic origin.

Region-specificity with associations to sympathetic and parasympathetic activations have also been reported [66]. Sympathetic activations are more associated with regions pertaining to executive and salience networks, i.e., anterior insula, anterior cingulate cortex, and further hubs in the prefrontal cortex; while parasympathetic activations are more associated with regions in the default mode network, i.e., posterior cingulate cortex, precuneus, angular gyrus, hippocampal formation. However, those regions represent a trend and should not be considered as sympathetic- or parasympathetic-exclusive structures *per-se*.

## IV. APPROACHES FOR QUANTIFYING COUPLING BETWEEN BRAIN-HEART TIME SERIES

Pairwise methods measuring the statistical association between variables have been exploited to explore brain-heart interactions through examination of the relationship between time series representative of the brain and cardiac oscillations. The adopted techniques span from conventional linear correlation methods [73], [74], [75] to frequency approaches including the cross-spectrum and the spectral coherence, also combined with information-theoretic methods [76], [77], and other nonlinear interdependence measures [27], [78], [79], [80]. With an approach as simple as the correlation function, analyses on source-reconstructed EEG signals have validated some of the findings from earlier neuroimaging studies [81], showing that the insula, amygdala, hippocampus, anterior and mid-cingulate cortices are involved in autonomic changes. Nevertheless, it should be remarked that a crucial aspect prior the application of any of such measures is the extraction of relevant variables from the brain and cardiac signals at hand, typically in the form of time series that capture synchronous information about brain and cardiovascular oscillations. This is usually done by building time series that map the dynamics of EEG oscillations computed via spectral analysis, and correlating them with HRV expressed by the series of the cardiac interbeat intervals at the scale of ~1 sec [26], [73], [74], [82], or with series mapping the sympathetic or parasympathetic component of HRV at longer time scales of ~1 min [24], [28], [80], [83]. On the other hand, alternative approaches looking directly at the cross-spectrum or spectral coherence between electrocardiogram and brain activity have been proposed [84], [85], but it is important to note that these approaches may not necessarily capture functional coupling. Instead, they often quantify isoelectric properties shared by the brain and the heart, which can sometimes arise as mere artifacts. Therefore, coherence analysis, particularly concerning HRV features [82], [83], should be prioritized to better understand the functional aspects of brain-heart coupling.

As an alternative to correlation and spectral coherence-based measures, model-free coupling and synchronization measures have been proposed to detect nonlinear dependencies between pairs of signals. In particular, the phase-space approach of synchronization likelihood [79] was proposed as a brain-heart coupling measure and tested on sleep EEG, showing a close relationship between a broad part of the EEG spectrum and high frequency HRV, being specially prominent in the delta-alpha range [80]. Further methodologies, such as Joint Symbolic Dynamics, detect patterns emerging from the interactions between two time series by coarse graining the series into sequences of symbols [86]. This approach was tested on patients suffering schizophrenia, showing insights into the effects of the anti-psychotic medication on the relationships between HRV, baroreflex and cortical dynamics [82], [86].

Overall, in the context of physiological couplings and the presence of non-linear dynamics, the most principled approaches are those framed in the field of information dynamics [87]. Understanding the intricate dynamics of information exchange between variables over time is crucial in various fields, particularly in studying complex systems like the nervous system. Formal quantification of information has become a basis in unraveling the complexities of information processing within physiological systems. The methodologies aiming to quantify information measures include mutual information, joint entropy, and instantaneous information shared between processes. Estimation of these measures often involves discretization of random variables through methods such as uniform quantization or rank ordering [88]. The Maximal Information Coefficient is a method proposed to quantify linear and non-linear correlations between two time series [78]. The computation is based on the mutual information between two time series, normalized by the minimum joint entropy. Among the method's advantages, it does not require symbolic transformations. Furthermore, the method may capture non-linear relationships, as it considers the similarities between two time series regardless their related distributions. This method has been tested in emotion elicitation studies, which revealed insights into the brain-heart dynamics associated to arousal in emotions [27].

Among the plethora of methodologies available, the combination of cross-spectral and information-theoretic approaches stands out as a promising tool for analyzing brain-heart interactions [76], [77]. Whereas spectral analysis provides a frequency-specific lens through which to examine the interactions between multiple time series, multivariate information measures allow the detection of information exchanges that may not be discernible through traditional time-domain analyses or spectral measures alone.

While traditional methods assume stationarity, physiological systems often exhibit non-stationary behavior. To address this limitation, time-varying approaches have been proposed [89]. These methods enable, for instance, the estimation of information storage at each time instant [90], capturing both abrupt and gradual changes in stored information over time. By applying these techniques to study brain-heart interactions, distinct variations can be unraveled in the variability of time-varying information storage across different phases of the cardiac cycle. This highlights the importance of considering non-stationarity in understanding dynamic processes within physiological systems.

Further methods exists to study linear interactions in the time and frequency domains as well as synchronizations and non-linear interdependences between pairs of time series [91], [92], [93], [94], [95], [96], [97]. However, most of these methods are derived from dynamical systems theory, in which the signals are used to reconstruct the underlying states of a latent dynamical system at every time. Although these methods can capture nonlinear couplings, they require a large amount of data to provide robust and unbiased estimators, and extremely sensitive to artefacts and nonstationary trends.

## V. CAUSALITY IN BRAIN-HEART INTERPLAY: ESTIMATION OF BOTTOM-UP AND TOP-DOWN INFORMATION FLOW

Brain connectivity measures [98] and methods investigating causal interactions in physiological signals [91], [99] are potential candidates to unveil brain-heart interactions. Causality holds significant relevance due to increasing evidence indicating a higher incidence of certain brain conditions in the presence of cardiovascular conditions, and vice versa. The bidirectional brain-heart relationship underscores the importance of understanding the causal pathways between the cardiovascular system and brain health. Identifying causal links can inform preventive strategies and interventions aimed at mitigating the risk and progression of both cardiovascular and neurological disorders. Existing tools aim to uncover the interactions within systems composed of multiple components. Key insights lie in discerning coupling direction, strength, and occurring time lags. Most used approaches rely on Granger-causality-based and entropy-based techniques quantifying the directed information transfer between signals and implemented via linear model-based or nonlinear model-free estimators [99], mutual nonlinear predictions detecting asymmetric relations in pairs of signals [91], [100], [101], and synthetic causal models of the underlying generative neural dynamics, among other connectivity measures [102].

Granger causality is a statistical method that aims to determine whether a time series is useful in forecasting another [103]. Therefore, Granger causality (GC) measures are candidate tools for assessing directional interactions between time series. In brief, the method is performed considering a N-dimensional stochastic process X = [$X_1$, …, $X_N$] and implementing a prediction model to estimate the information transferred from the scalar process $X_i$ to the process $X_j$ (i,j ∈ 1,...,N, i ≠ j); the analysis is performed either discarding (bivariate GC) or taking into account (multivariate or conditional GC) the remaining N-2 processes. In either case, causality estimation is based on comparing the extent to which the knowledge of the past states of the putative driver $X_i$ improves the prediction (by reducing the error of the model) of the present state of the target $X_j$, above the extent to which the driver is predicted by its own past states (and of the past states of the other processes in the multivariate case). While traditional GC approaches rely on linear regression, nonlinear prediction models can also be adopted [104], [105]. Granger causality has been primarily used to describe brain network connectivity [106], [107] and cardiovascular interactions [108],

[109], but also to gather brain-heart interactions. For instance, Duggento and colleagues revealed that some regions previously described as correlated with autonomic dynamics are actually associated with brain-to-heart neural control [72], [110]. Faes et al. [28] characterized the topology of brain-heart interaction networks during sleep using GC, highlighting bidirectional communications between the cardiac parasympathetic variability and the beta EEG activity and unidirectional brain-to-heart interactions when slower brain waves are considered [28]. Further evidence on Granger causality applied to EEG showed that brain-to-heart coupling increases in the left hemisphere for positive emotional valence and in the right hemisphere for negative valence, as gathered from prefrontal, somatosensory, and posterior cortices [111]. Clinical evidence has been provided for the cases of sleep apneas and epilepsy. During sleep recordings, GC revealed differences between healthy controls and patients suffering obstructive sleep apnea, where bidirectional brain-heart coupling in the lower frequency ranges could distinguish between the participants' groups [112]; an impaired brain-to-heart communication during severe sleep apnea-hypopnea syndrome was detected using GC computed across whole-night recordings [113], also showing the potential of long-term ventilation therapy to recover the physiological brain-heart interaction patterns. In epilepsy, GC revealed a dominance of the brain-to-heart causality, over the heart-to-brain counterpart, suggesting that the central control on autonomic dynamics during the ictal phase of the seizures [114].

The concept of statistical causality has been formalized also in terms of entropy-based methods such as the Transfer Entropy (TE), which provides a model-free probabilistic tool to assess the information transfer between two time series [115]. While standard mutual information approaches fail in distinguishing the directed information exchange between processes, information transfer estimates can distinguish the driving and responding elements, and detect potential asymmetries on the interactions [115], [116]. TE has been proposed as an alternative measure of effective connectivity [117], used as well to describe brain-heart interplay, which has been tested in sleep and schizophrenia. In sleep EEG, TE revealed that the beta power conveys the largest amount of bidirectional brain-heart information flow across different sleep stages, being weaker in the transitions from light sleep to deep sleep and to REM sleep [24]; a direct comparison between linear GC and nonlinear TE evidenced the role of nonlinear correlations in driving brain-heart interactions during sleep [28]. In schizophrenia, transfer entropy revealed stronger heart-to-brain influences, as compared to healthy controls [26]. A normalized version of the TE, estimated via non-uniform embedding [116] between the time series of HRV and EEG complexity, was employed as well by Yu and colleagues [118], who revealed the existence of unidirectional effects of the cardiac period length on the irregularity of the brain waves in the resting and mental stress states. A similar approach was employed to distinguish between physiological changes induced by internally-driven attention, linked with short-term memory assessment, and externally-driven attention, associated with automatic and transient responses to external stimuli; the findings revealed that heart-to-brain information flow increased, while the brain-to-heart

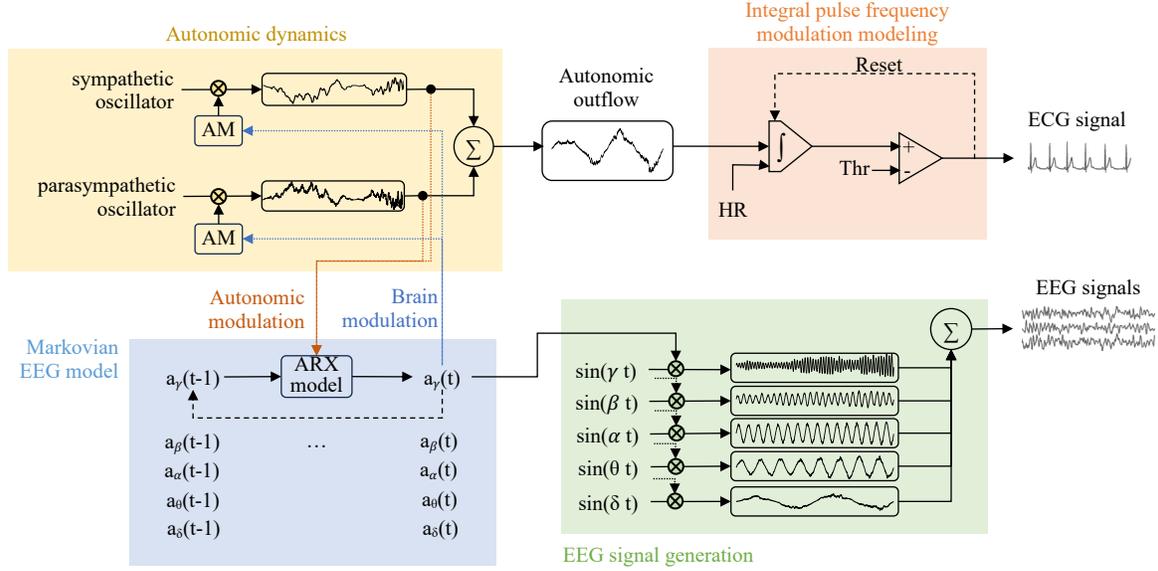

Fig. 4. Modeling of bidirectional brain-heart interaction through block diagrams of the coupled heartbeat and brain signal generation systems. The heartbeats' generation in the sinoatrial node is modeled as an integrate-and-fire model (red block), namely integral pulse frequency modulation model, which receives as an input the sum of sympathetic and parasympathetic inputs and the baseline heart rate (HR). The model generates the heartbeats as a train of pulses at each time the integration reaches a defined threshold (Thr). Autonomic dynamics (yellow block) are disentangled in the sympathetic and parasympathetic components, which are individually modeled as an oscillator whose amplitude is modulated (AM) on time, as a function of the changes in EEG power. In the brain part, EEG signals are modeled as the sum of five frequency bands (green block), typically, δ: 0-4 Hz, θ: 5-8 Hz, α: 9-12 Hz, β=13-30 Hz, γ: 31-50 Hz, whose powers (a_F; with F: δ, θ, α, β, γ) are individually modeled (blue block) as an autoregressive process that receives autonomic modulations as an external term (ARX model).

flow decreased, during externally-driven attention compared to internally-driven attention [119].

Alternative approaches for the estimation of directional coupling between pairs of time series make use of the concept of cross-predictability, whereby embedding vectors from one series are used to predict future states of the other. These approaches lay their ground on the theory of dynamical systems and are based on the concept of state-space correspondence [101], whereby it is assumed that it should be possible to cross-map between the variables observed from a system and extract predictability measures out of such cross-mapping. The most popular method in this context is the Convergent Cross-Mapping (CCM) [91], a statistical tool for cross-prediction which exploits the idea that the reconstructed states from a responding signal can be used to cross-map the driver signals. Convergent Cross-Mapping has been used to study brain-heart interactions [25], with particular evidence on epilepsy [120], [121], [122], [123]. Interestingly, although the underlying concept and assumptions are different, CCM and GC yielded overlapping results on the analysis of time series of HRV features and the envelopes of delta and alpha EEG activity [114], [123]. These consistent findings suggest that cortical oscillations drive the autonomic activity before, during and after the development of epileptic seizures.

Synthetic data generation are a new framework that aims to model causality and directionality of the neural modulations following a logic of generative neural dynamics (Fig. 4). These models assess the bidirectional modulations between EEG oscillations (at a given frequency band) and heartbeat dynamics time series [29], [30], [124]. The estimates of brain-to-heart interplay are quantified through a model able to generate synthetic heartbeats, which are represented as a integrate-and-fire model. These models of synthetic heartbeats are referred as Integral Pulse Frequency Modulation models [125], [126], [127], [128], [129], [130]. The heart-to-brain interplay is usually quantified through a model based on the generation of synthetic EEG series using an adaptative Markov process on brain power series [131]. The model estimates the ascending modulations from the heart to the brain using least squares in a first-order auto-regressive process, in which the Markovian neural activity generation uses its previous neural activity and the current heartbeat dynamics as inputs. This approach offers a time-resolved estimation of bidirectional brain-heart interactions, which has been used to model the physiological dynamics in emotions, showing that ascending cardiac inputs modulate brain dynamics in different contexts of arousal [30], [132], [133]. This modeling has also been tested in clinical conditions, including mood disorders [134] and coma patients [135].

## VI. BRAIN-HEART HIGHER-ORDER INTERACTIONS

Complex systems often exhibit interactions among multiple components that go beyond simple pairwise connections, involving higher-order interactions among three or more nodes [136]. These higher-order interactions can significantly impact collective network behavior but, are often overlooked in traditional analyses. To address this gap, methodologies have been developed to characterize both pairwise and higher-order interactions among multivariate time series (Fig. 5a), with a focus on assessing the equilibrium between redundant and synergistic information. In the realm of information theory, several distinct types of information are recognized, each offering valuable perspectives on system dynamics.

Synergistic, redundant, and unique information are among these types [137]. Synergy arises from collective statistical interactions within a network which cannot be inferred when the sources of information are considered in isolation; as such, synergy amplifies the efficiency of information exchange by leveraging interactions among multiple system elements. Redundancy, on the other hand, encapsulates information that is conveyed equally by more sources; it ensures system robustness but at the expense of not fully utilizing the available information capacity. These characterizations offer a more detailed understanding of how information flows and is used within complex systems.

Two main methodological frames have been conceptualized for the analysis of higher-order interactions. The first is based on the Shannon theory of information, which captures the balance between redundant and synergistic information among groups of three or more variables via measures like the so-called interaction information [138] and its generalizations (prominently, the so-called O-information [139]). The framework of partial information decomposition [90] is more powerful as it provides separate measures of synergy and redundancy, but also more complicated because its unequivocal formulation requires to go beyond classical information theory [140]. Though being still under active development, these frameworks are gaining strong relevance for the analysis of multivariate biological systems, for instance in the brain and in the cardiovascular system where redundancies and synergies have been found to play distinct crucial roles in explaining the mechanisms that govern robust and flexible physiological regulation [137], [141], [142]. Therefore, it is expected that higher-order interaction frameworks will become soon pervasive in the analysis of brain-heart communications sustained by multi-layer networks with activities deployed across multiple spatial and temporal scales (Fig. 5d).

While some studies have started analyzing the spatial distribution of the brain activity associated with cardiac dynamics in terms of high-order brain-heart interactions [74], [114], the investigation of multi-scale behavior has been limited by the fact that the theoretical frameworks have been formalized with focus on random variables, thus with no explicit account of temporal correlations. However, the recent introduction of a dynamic framework for the analysis in networks of random processes, formalized via measures of entropy and mutual information rates [76], [143], [144], has introduced an approach to assess higher-order interactions in rhythmic processes with rich oscillatory content. The O-information rate, which assesses the equilibrium between redundancy and synergy [143], exploits spectral representations of vector autoregressive and state space models to assess interactions among groups of processes, both in specific frequency bands and in the time domain after whole-band integration. It allows to highlight redundant and synergistic interactions emerging at specific frequencies, offering insights not detectable using traditional time-domain measures. One can relate the O-information to pairwise measures of dynamic coupling like the spectral coherence [144], decompose it into measures quantifying Granger causality and instantaneous influences in different frequency bands [143], or examine its gradients to derive low-order descriptors offering insights into the individual contributions of variables in shaping high-order informational circuits [145].

Further statistical inference methodologies can be embedded in these higher-order interactions frameworks [146], to characterize functional links within physiological networks. Validation on theoretical and numerical simulated networks demonstrated its ability to represent higher-order interactions, but also to detect cascades, by dynamically identifying drivers and targets within the networks. These approaches aim at further describing the hierarchical dynamics within the system, allowing the evaluation of dynamic networks depicted by multivariate time series [144], offering versatility and scalability for exploring interactions beyond pairwise connections.

## VII. Cardiac-related brain network dynamics

Most of previous studies have predominantly focused on the interaction between specific brain or scalp regions and heartbeat dynamics, disregarding the dynamic nature of brain networks and their role in numerous neural functions [73], [143], [147], [148]. In line with this, there have been proposals for frameworks to study brain-heart interactions that explore the relationship between ongoing brain network organization and cardiac oscillations. Some fMRI studies have explored the relationship between HRV and the connectivity in certain brain regions [149], [150], [151]. However, these approaches rely on the definition of a seed, which is typically defined as the main nodes in the central autonomic network. In a more agnostic manner, these frameworks can be extended to the identification of the brain networks associated to changes in cardiac dynamics in certain conditions. In a recent study, authors examine the interplay between pairwise brain connectivity and cardiac dynamics (Fig. 5c). This framework explores the relationship between triads by quantifying the coupling between the pairwise brain region connectivity and the cardiac dynamics, with the ultimate goal of identifying all the links associated to a network that is formed together with cardiac dynamics under different conditions [152].

Another related framework provides biomarkers related to large-scale brain-heart interaction by quantifying the intricate dynamics between global brain activity and cardiac dynamics [153] (Fig. 5b). This framework showcases how the study of brain-heart interactions can be approached in various conditions where global neural dynamics are not fully understood by solely examining the dynamics of specific brain regions. It delves into the variations in global network dynamics, focusing on parameters such as efficiency, clustering, modularity, and assortativity within brain connectivity matrices [154]. The aim of this framework is to provide a holistic quantification of global dynamics and their relationships with the fluctuations in cardiac sympathetic-vagal dynamics [153].

As a proof-of-concept, it was proposed to approach brain-other organs interactions through frameworks of multi-layer networks[155] (Fig. 5d). The multilayer structure provides a comprehensive framework for understanding, for instance, complex interactions by incorporating detailed structural and functional information across multiple levels [156]. Multilayer networks analysis offers a means to study the human brain's diverse functional layers, enabling the potential integration of brain-heart interactions. This framework may also allow the modeling of the interactions at different scales, from molecular to systemic mechanisms [14] at different layers. Importantly, multilayer analysis may contribute to transcend from brain-centered analyses, fostering a holistic understanding of brain-other organ interactions. Operationalizing multilayer definitions depends on the specific phenomena being modeled, offering flexibility in adapting to various research contexts. By integrating empirical evidence from brain-heart interplay, multilayer frameworks holds promise as an integrative framework for advancing our understanding of complex systems.

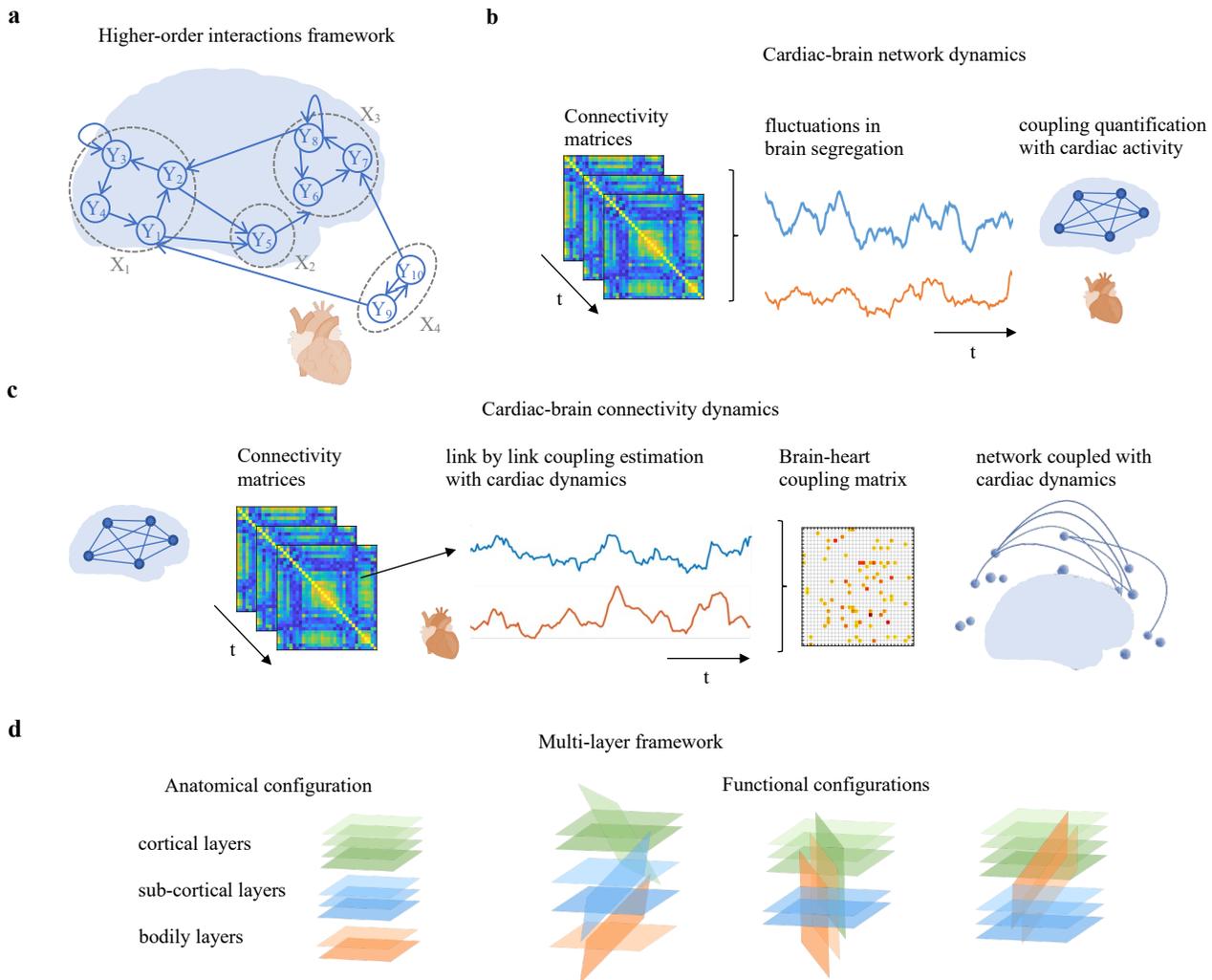

Fig. 5. Frameworks that relate to higher-order brain-heart interaction. (a) Brain-heart interactions can be accounted as complex systems' analysis with multi-node interplay. (b) Cardiac-brain network dynamics aims at quantifying the relationship of network measures, such as integration and segregation, and parallel changes in cardiac dynamics. (c) Cardiac-brain connectivity dynamics aims at quantifying the relationship between brain connectivity and cardiac dynamics, by identifying the individual brain links that change in parallel to cardiac dynamics, to ultimately identify the whole network associated to these changes. (d) Multi-layer frameworks aim at studying brain-heart interactions by modeling the different nodes within pre-defined layers and re-refined ones as per their functional relationships.

## VIII. Methodological applicability

The applicability of methods for analyzing brain-heart interactions depends on various factors, with the most significant being the amount of available data, prior physiological knowledge, and underlying hypotheses regarding time, frequency, and regional dimensions. Below, we outline some of these factors that need to be considered, including aspects related to physiological causality analysis [102], which are not exclusive to brain-heart interactions.

### A. Limited amount of data

When data is scarce, visualizing brain-heart interactions graphically can be more valuable. Symbolic transformations of the data are ad-hoc methods commonly used in such instances. These approaches may prove suitable due to their robustness against noise and potential accounting for nonlinearities. They often prioritize pairwise comparisons over multivariate analyses.

Additionally, the analysis of HRV is limited by the length of the recording. Short-term recordings, typically one minute or less, may not capture the full range of HRV fluctuations of interest, and are more susceptible to transient artifacts and anomalies [157]. Therefore, balancing the duration of recordings as a function of the expected physiological outcomes is necessary to achieve reliable and representative HRV measurements.

### B. Large datasets

In large datasets containing rich information from multiple brain regions, model-free methods offer computationally tractable estimations and can facilitate an agnostic search for hierarchical dynamics. Within this framework, phase space methods provide a model-free approach to detect such dynamics, with possibility to extend for multivariate analyses.

### C. System agnostic analyses

Without prior knowledge of the systems studied, coupling measures drawn from information theory offer versatility, as they are less likely to overlook nonlinear couplings. Information theory-based method can be further extended to have notions of causality and multivariate interactions. However, these analyses have to be re-considered in cases of limited amount of data, given the computational resources needed, for instance, to estimate probability distributions.

### D. Prior knowledge available

With prior knowledge at hand, measures and models investigating causality are likely to offer a more informative framework. Some of these approaches are easily adaptable to the available data and can be applied directly or in transformed and multivariate analyses. The expected complexity of these methods should be balanced with the available data. For instance, if multiple auto-regressive processes need to be conducted, detecting only linear interactions may suffice for the analysis needs.

### E. Region-agnostic analyses

When the goal is to comprehend global neural dynamics without prior hypotheses regarding specific brain regions involved, network-based analyses may be suitable. This is particularly applicable in cases where there is a high level of uncertainty in the data due to population heterogeneity or neural damage preventing analysis in the same region for all subjects. Methods based on global network dynamics offer characterizations that are independent of specific regions, enabling comparisons of global dynamics across the subject's heterogeneity. However, these approaches may necessitate control measures to ensure that the effects observed are not attributable to differences in the number of nodes or network densities.

## IX. METHODOLOGICAL CHALLENGES

### A. Improving specificity of the brain regions involved in brain-heart interactions

Current data-driven methods for inferring and analyzing complex networks involve constructing a model from observed time series, where nodes represent units in the system and edges/hyperedges signify functional pairwise/higher-order dependencies between these units. However, most of the models are constructed from scalp recorded EEG signals, often overlooking subcortical dynamics. Future research should aim at identifying the specific cortical sources and their connectivity underlying the sensor/electrode level interactions [158]. This challenge can be approached by adapting physiological models using scalp EEG recorded signals to cortical source-reconstructed signals [81], but also in adaptations for fMRI recordings. These efforts will include modeling the relationships between electrophysiological activity and metabolic activity, by accounting the different confounding factors that appear when measuring in parallel various signals with different generative nature.

### B. Improving the estimation of the directionality of the interactions between the brain and heartbeat dynamics

On the one hand, distinguishing genuine interactions from mere correlations between brain and cardiac signals posits a utmost challenge. This require emphasizing the need for methodologies capable of uncovering true causation and functional relationships. In this line, methods accounting for the physiological priors, such as generative modeling [30], [124], [132], could diminish the quantification of spurious correlations.

On the other hand, deciphering the physiological dynamics may highly rely on understanding which system influences the other in specific conditions. While time-resolved estimations enable the observation of dynamic interactions within closed-loop physiological circuits, those works rely on pairwise measures to assess functional dependencies. However, these measures may not fully capture the interactions within complex systems that often exhibit collective behaviors at different hierarchical levels, involving more than two network nodes. To address this limitation, methods may explore multivariate approaches like conditional causality measures and higher-order interactions among multiple time series. In this direction, the incorporation of strategies to uncover hierarchies of the interactions would enhance the causality estimation [144].

### C. Achieving high time-resolution in the multiscale and time-resolved estimation of brain-heart interactions

Given that physiological dynamics at brain and heart level occur at different time scale, the unfolding of the complex interplay between brain and cardiac activities at high sampling rate posits different challenges, based on the brain recording modality. These challenges can be primarily addressed by optimizing time-frequency analyses by incorporating state-of-the art solutions for better time resolution, such as Wavelet transforms and smoothed Wigner-Ville distributions [159], [160]. In this line, development in time-resolved estimations may provide sufficient information for real-time applications, such as brain-computer interfaces and neurofeedback. Another crucial development, related to the different scales of oscillations typically observed between brain and heart rhythms, is the development of multi-scale and cross-scale markers allowing to explore new modes of brain-heart interaction.

### D. Overlooking the complex network nature of brain-heart interactions

While existing information-theoretic measures provide valuable insights, a limitation is their characterization of system dynamics with a single value, overlooking the rich oscillatory content inherent in complex network time series. For instance, brain-heart interactions involve rhythms in different frequency bands with varying physiological significance, requiring an approach that connects spectral representation with higher-order interactions. In response to this, some frameworks account for the time- and frequency-domain analysis of higher-order interactions in multivariate stochastic processes mapping network system activity [143]. Upon these frameworks, the incorporation of multivariate decompositions may provide a better understanding of complex network dynamics, particularly in capturing the diverse nature of higher-order interactions across different bodily rhythms.

### E. Overlooking the hierarchy of neural oscillations

While multiple nodes may interact dynamically over time, a characterization of the hierarchical architecture of network dynamics remains challenging. In this line, the identification of the leading nodes of a complex system may be relevant for targeted treatments using neuromodulation, pharmacologically or brain stimulation techniques. Network science techniques can be employed to further describe these interactions, exploring the level of brain network controllability from visceral or peripheral bodily inputs, studying the nodes in charge of network integration and segregation in relation to peripheral bodily inputs, and employing various approaches to estimate causality and directionality of those measures among multiple-node signals.

## X. TRANSLATIONAL PERSPECTIVES

### A. Mental health and neuromodulation treatments

The relationship between mental health and the bodily state has been paramount. Recognizing dysfunctions in interoception has become increasingly important in understanding various mental health conditions, including anxiety disorders, mood disorders, eating disorders, addictive disorders, and somatic symptom disorders [161]. Our understanding of the complex relationship between mental health and interoception has primarily been informed by behavioral evidence [162], yet a physiological explanation of affective states and associated disorders remains elusive [163].

Research into clinical and subclinical depression has predominantly centered on examining the brain dynamics of individuals exhibiting depressed mood symptoms. However, systematic analysis show that brain imaging-based biomarkers cannot identify depression at individual level [164]. Moreover, evidence has shown that depression extends beyond a brain-exclusive disorder; it is intricately linked to cardiovascular conditions [165]. For instance, mood disorders are linked to an increased risk and more unfavorable prognosis of coronary heart disease [166]. Conversely, individuals with cardiac pathologies exhibit a higher prevalence of depression and depressive symptoms when compared to the general population [167], [168]. Recent research on brain-heart interactions suggests that depressed mood is associated with an intensified control over slow HRV changes [134] or reduced control over the fast ones [169]. Similarly, some of those dynamics have been shown in anxiety as well [170]. These preliminary results suggest that research into brain-heart interactions holds immense promise for advancing the development of improved biomarkers crucial for the detection, prevention, and stratification of mental health conditions.

Beyond diagnosis and disorders' characterization, some treatments for depression include the use of transcranial magnetic stimulation (TMS), which uses magnetic fields to stimulate brain regions. However, the physiological mechanisms are not fully understood. From a systems perspective, TMS is believed to induce neuroplastic changes in the brain, promoting the formation of new neural connections, but also to modulate the excitability of neural circuits in the prefrontal cortex, which may contribute to restoring more balanced and healthy neural functioning [171]. However, several factors have been found or hypothesized to alter TMS effectiveness in treating depression; for instance, the specific TMS treatment protocol (location, frequency, intensity, and duration of sessions) can impact its effectiveness, or the inter-subject variability in their pathology phenotype and their TMS responses. Recent work introduces the concept of a brain-heart network, which intersects with the functional nodes of the previously described depression network [171]. Neuromodulation studies using TMS typically trigger key nodes within this network, specifically, modulations to the dorsolateral prefrontal cortex and the anterior cingulate cortex, which have subsequent impact on cardiac dynamics. This evidence emphasizes the significance of incorporating brain-heart interplay measurements in human depression treatments, especially those involving neuromodulation. Further developments on this can target the heterogeneity of the patients' outcomes using neuromodulation, and therefore, allowing the potential development of personalized therapies for depression. In this line, Neuro-Cardiac-Guided TMS treatment has emerged motivated by this brain-heart network [172], which enhances the precision on the TMS location. While a thorough delineation of the physiological dynamics is still pending. The frameworks of brain-heart interplay present a compelling approach to tackle the difficulties of neuromodulation in depression, which may provide biomarkers capable of stratifying patients based on their anticipated outcomes.

### B. Neurodegeneration, stroke, and rehabilitation engineering

Autonomic dysfunction under neurodegeneration can involve various bodily systems, including those generating the cardiovascular dynamics [173], [174]. Further research in different neurodegenerative conditions has suggested a disruption in the awareness of one's own heartbeats, as measured from cardiac interoception tasks [53], [54], [175], [176], [177], suggesting a disruption in the communication between the brain and the heart. However, only recently brain-heart interactions have been assessed in these conditions, for instance to characterize autonomic dysfunctions [178], orthostatic hypotension [179], or dopaminergic therapy effects on motor symptoms [152], [153]. On the other hand, brain damage caused by stroke can lead extensive changes in the nervous system as well. Abundant evidence exists on the brain-heart effects caused by stroke, from molecular to systemic changes [14].

One of the main challenges in these conditions is the recovery of sensorimotor functions. Recently, brain-heart interactions have shown a close relationship with different aspects of responsiveness, decision-making and motor functions [40], [41], [42], [43], [44], [45], [180]. In particular, this knowledge can be embedded in brain-computer interfaces (BCIs), as they hold promise in the restoration of lost sensorimotor abilities after suffering brain damage from conditions such as Parkinson's disease, stroke and multiple sclerosis [181]. However, their effectiveness varies because BCIs typically need to be customized for each patient [182], [183]. For this, the development of objective markers for monitoring task performance, learning, and progress remains one of the main challenges in BCIs [182], [183]. The study of brain-heart interactions in motor imagery and BCIs has emerged only recently [184], [185]. These studies show that biomarkers based on brain-heart interactions hold promise in identifying distinct couplings with respect to cognitive and sensorimotor synergies. Therefore, the understanding of the specific contributions of cardiac dynamics can further enlighten the rehabilitation of sensorimotor abilities, by either facilitating the relearning of movements and enhancing functional recovery, for instance, by enabling patients to control on screen commands, robotic exoskeletons, or prosthetic limbs [186].

Neural damage resulting from neurodegeneration or stroke often extends beyond specific brain regions, impacting various parts of the nervous system. In contrast to being solely localized to one area, this widespread pathology underscores the importance of exploring brain-heart interactions. Such investigations offer valuable frameworks for comprehending the physiopathology of these diseases and designing rehabilitation strategies, particularly those leveraging BCIs.

### C. Severe brain damage, consciousness, and neuroprognostication

In clinical practice, standard consciousness assessment after severe brain damage relies on characterizing bedside responsiveness [187]. Therefore, presence of consciousness is often associated with the detection of non-reflex behavior. However, the challenge arises on the patients' high heterogeneity in their clinical phenotype, which may be translated in different sensorimotor impairments and fluctuations in vigilance, leading to a high rate of misdiagnosis [188], [189]. The research of an accurate consciousness diagnosis based on behavior, the exploration of neuroimaging and electrophysiology techniques to reduce the misdiagnosis rate has been part of extensive consciousness research [190]. One of the explored approaches is based on the variability of the neural responses to heartbeats at the bedside, indicative of the relationship between the presence of consciousness and a healthier brain-heart connection [52]. These responses to heartbeats were found to complement other EEG-based markers of consciousness [191], and to be more complex and more segregated through the scalp, as a function of the level of consciousness [61]. Further evidence revealed that cardiac inputs in the brain seem to participate in the conscious processing of ongoing exteroceptive information, which also appears as a signature of consciousness in these patients [192], [193].

Exploring brain-heart interactions can provide valuable insights into the physiological condition following severe brain damage. This approach holds promise for identifying biomarkers that could aid in addressing the ongoing challenges faced in the clinical practice on these patients. For instance, prognostication remains challenging due to our limited understanding of the multisystem physiological implications caused by severe brain damage. In this direction, patients with severe post-cardiac arrest brain injury were found to display bidirectional brain-heart interactions that scale with the severity of the brain injury and with patients' neurological outcome at 3 months [135].

Thus, the field of research focusing on brain-heart interactions offers a promising avenue for unraveling the complexities of physiology following severe brain damage. By investigating this relationship, diagnostic and prognostic biomarkers can be identified to provide valuable insights into the clinical phenotype of these patients. Ultimately, such advancements have the potential to revolutionize the way we understand and manage critical care monitoring, offering personalized approaches tailored to their specific needs and conditions.

## XI. CONCLUSION

There is an abundance of coupling measures that can be exploited for the analysis of brain-heart interplay, each with its own set of advantages and drawbacks. Beyond an overview of the methods, we aimed to provide some conclusive remarks and guidance on when to explore specific frames providing coupling measures, based on the study objective and the data available. In doing this, we highlighted the key methodological challenges present in current approaches to measuring and modeling brain-heart interactions. Addressing these challenges will undoubtedly enhance our understanding of the physiological mechanisms underlying various neural functions, including interactions between the brain and other organs. Our outlines also offer insights into a research agenda aimed at advancing methods for accurately estimating brain-heart interactions.

The ongoing evidence is uncovering non-linear, complex, and bidirectional communications between brain and heart dynamics. Further developments on these methodologies will contribute to a better understanding of the physiological dynamics involved in regulation mechanisms, predicting coding, and cognitive functions.

Finally, we highlighted some significant advancements in understanding the physiopathology of diseases and their connections with brain-heart interactions. These advancements prompt new research avenues where brain-heart interactions play a crucial role in understanding diseases, transforming them into diagnostic tools. Additionally, they offer insights into prognostic tools, treatment evaluation, and the design of personalized and targeted interventions.

## REFERENCES

[1] R. Aschenbrenner and G. Bodechtel, "Über Ekg.-Veränderungen bei Hirntumorkranken," *Klin Wochenschr*, vol. 17, no. 9, pp. 298–302, Feb. 1938, doi: 10.1007/BF01778563.

[2] A. Silvani, G. Calandra-Buonaura, R. A. L. Dampney, and P. Cortelli, "Brain-heart interactions: physiology and clinical implications," *Philos Trans A Math Phys Eng Sci*, vol. 374, no. 2067, May 2016, doi: 10.1098/rsta.2015.0181.

[3] S. Pyner, "The paraventricular nucleus and heart failure," *Experimental Physiology*, vol. 99, no. 2, pp. 332–339, 2014, doi: 10.1113/expphysiol.2013.072678.

[4] W. G. Chen *et al.*, "The Emerging Science of Interoception: Sensing, Integrating, Interpreting, and Regulating Signals within the Self," *Trends in Neurosciences*, vol. 44, no. 1, pp. 3–16, Jan. 2021, doi: 10.1016/j.tins.2020.10.007.

[5] K. S. Quigley, S. Kanoski, W. M. Grill, L. F. Barrett, and M. Tsakiris, "Functions of Interoception: From Energy Regulation to Experience of the Self," *Trends in Neurosciences*, vol. 44, no. 1, pp. 29–38, Jan. 2021, doi: 10.1016/j.tins.2020.09.008.

[6] H. D. Critchley and N. A. Harrison, "Visceral influences on brain and behavior," *Neuron*, vol. 77, no. 4, pp. 624–638, Feb. 2013, doi: 10.1016/j.neuron.2013.02.008.

[7] M. A. Samuels, "The Brain–Heart Connection," *Circulation*, vol. 116, no. 1, pp. 77–84, Jul. 2007, doi: 10.1161/CIRCULATIONAHA.106.678995.

[8] A. Jaggi *et al.*, "A structural heart-brain axis mediates the association between cardiovascular risk and cognitive function," *Imaging Neuroscience*, vol. 2, pp. 1–18, Jan. 2024, doi: 10.1162/imag_a_00063.

[9] B. Zhao *et al.*, "Heart-brain connections: Phenotypic and genetic insights from magnetic resonance images," *Science*, vol. 380, no. 6648, p. abn6598, Jun. 2023, doi: 10.1126/science.abn6598.

[10] J.-C. Jiang, K. Singh, L. K. Davis, N. R. Wray, and S. Shah, "Sex-Specific Association Between Genetic Risk of Psychiatric Disorders and Cardiovascular Diseases," Sep. 26, 2023, *medRxiv*. doi: 10.1101/2022.10.08.22280805.

[11] K. Gabisonia, M. Khan, and F. A. Recchia, "Extracellular vesicle-mediated bidirectional communication between heart and other organs," *Am J Physiol Heart Circ Physiol*, vol. 322, no. 5, pp. H769–H784, May 2022, doi: 10.1152/ajpheart.00659.2021.

[12] P. Lackner *et al.*, "Cellular Microparticles as a Marker for Cerebral Vasospasm in Spontaneous Subarachnoid Hemorrhage," *Stroke*, vol. 41, no. 10, pp. 2353–2357, Oct. 2010, doi: 10.1161/STROKEAHA.110.584995.

[13] Y. Yang and G. A. Rosenberg, "Blood–Brain Barrier Breakdown in Acute and Chronic Cerebrovascular Disease," *Stroke*, vol. 42, no. 11, pp. 3323–3328, Nov. 2011, doi: 10.1161/STROKEAHA.110.608257.

[14] Z. Chen, P. Venkat, D. Seyfried, M. Chopp, T. Yan, and J. Chen, "Brain–Heart Interaction," *Circulation Research*, vol. 121, no. 4, pp. 451–468, Aug. 2017, doi: 10.1161/CIRCRESAHA.117.311170.

[15] G. Long *et al.*, "Circulating miR-30a, miR-126 and let-7b as biomarker for ischemic stroke in humans," *BMC Neurology*, vol. 13, no. 1, p. 178, Nov. 2013, doi: 10.1186/1471-2377-13-178.

[16] M.-L. Liu and K. J. Williams, "Microvesicles: potential markers and mediators of endothelial dysfunction," *Curr Opin Endocrinol Diabetes Obes*, vol. 19, no. 2, pp. 121–127, Apr. 2012, doi: 10.1097/MED.0b013e32835057e9.

[17] X. J. Wei *et al.*, "Biological significance of miR-126 expression in atrial fibrillation and heart failure," *Braz J Med Biol Res*, vol. 48, pp. 983–989, Aug. 2015, doi: 10.1590/1414-431X20154590.

[18] L. Jammal Salameh, S. H. Bitzenhofer, I. L. Hanganu-Opatz, M. Dutschmann, and V. Egger, "Blood pressure pulsations modulate central neuronal activity via mechanosensitive ion channels," *Science*, vol. 383, no. 6682, p. eadk8511, Feb. 2024, doi: 10.1126/science.adk8511.

[19] W.-Z. Zeng *et al.*, "PIEZOs mediate neuronal sensing of blood pressure and the baroreceptor reflex," *Science*, vol. 362, no. 6413, pp. 464–467, Oct. 2018, doi: 10.1126/science.aau6324.

[20] G. Dirlich, L. Vogl, M. Plaschke, and F. Strian, "Cardiac field effects on the EEG," *Electroencephalography and Clinical Neurophysiology*, vol. 102, no. 4, pp. 307–315, Apr. 1997, doi: 10.1016/S0013-4694(96)96506-2.

[21] A. D. Craig, "How do you feel? Interoception: the sense of the physiological condition of the body," *Nat. Rev. Neurosci.*, vol. 3, no. 8, pp. 655–666, Aug. 2002, doi: 10.1038/nrn894.

[22] L. I. Skora, J. J. A. Livermore, and K. Roelofs, "The functional role of cardiac activity in perception and action," *Neuroscience & Biobehavioral Reviews*, vol. 137, p. 104655, Jun. 2022, doi: 10.1016/j.neubiorev.2022.104655.

[23] D. Azzalini, I. Rebollo, and C. Tallon-Baudry, "Visceral Signals Shape Brain Dynamics and Cognition," *Trends in Cognitive Sciences*, vol. 23, no. 6, pp. 488–509, Jun. 2019, doi: 10.1016/j.tics.2019.03.007.

[24] L. Faes, G. Nollo, F. Jurysta, and D. Marinazzo, "Information dynamics of brain–heart physiological networks during sleep," *New J. Phys.*, vol. 16, no. 10, p. 105005, Oct. 2014, doi: 10.1088/1367-2630/16/10/105005.

[25] K. Schiecke, A. Schumann, F. Benninger, M. Feucht, K.-J. Baer, and P. Schlattmann, "Brain-heart interactions considering complex physiological data: processing schemes for time-variant, frequency-dependent, topographical and statistical examination of directed interactions by

convergent cross mapping," *Physiol Meas*, vol. 40, no. 11, p. 114001, Dec. 2019, doi: 10.1088/1361-6579/ab5050.

[26] S. Schulz, J. Haueisen, K.-J. Bär, and A. Voss, "Altered Causal Coupling Pathways within the Central-Autonomic-Network in Patients Suffering from Schizophrenia," *Entropy*, vol. 21, no. 8, Art. no. 8, Aug. 2019, doi: 10.3390/e21080733.

[27] G. Valenza *et al.*, "Combining electroencephalographic activity and instantaneous heart rate for assessing brain–heart dynamics during visual emotional elicitation in healthy subjects," *Philosophical Transactions of the Royal Society A: Mathematical, Physical and Engineering Sciences*, vol. 374, no. 2067, p. 20150176, May 2016, doi: 10.1098/rsta.2015.0176.

[28] L. Faes, D. Marinazzo, F. Jurysta, and G. Nollo, "Linear and non-linear brain-heart and brain-brain interactions during sleep," *Physiol Meas*, vol. 36, no. 4, pp. 683–698, Apr. 2015, doi: 10.1088/0967-3334/36/4/683.

[29] V. Catrambone, A. Greco, N. Vanello, E. P. Scilingo, and G. Valenza, "Time-Resolved Directional Brain-Heart Interplay Measurement Through Synthetic Data Generation Models," *Ann Biomed Eng*, vol. 47, no. 6, pp. 1479–1489, Jun. 2019, doi: 10.1007/s10439-019-02251-y.

[30] D. Candia-Rivera, V. Catrambone, R. Barbieri, and G. Valenza, "Functional assessment of bidirectional cortical and peripheral neural control on heartbeat dynamics: a brain-heart study on thermal stress," *NeuroImage*, vol. 251, p. 119023, Feb. 2022, doi: 10.1016/j.neuroimage.2022.119023.

[31] R. Salomon *et al.*, "The insula mediates access to awareness of visual stimuli presented synchronously to the heartbeat," *Journal of Neuroscience*, vol. 36, no. 18, pp. 5115–5127, 2016, doi: 10.1523/JNEUROSCI.4262-15.2016.

[32] A. Schulz *et al.*, "Cardiac cycle phases affect auditory-evoked potentials, startle eye blink and pre-motor reaction times in response to acoustic startle stimuli," *International Journal of Psychophysiology*, vol. 157, pp. 70–81, Nov. 2020, doi: 10.1016/j.ijpsycho.2020.08.005.

[33] L. Edwards, C. Ring, D. McIntyre, J. B. Winer, and U. Martin, "Sensory detection thresholds are modulated across the cardiac cycle: Evidence that cutaneous sensibility is greatest for systolic stimulation," *Psychophysiology*, vol. 46, no. 2, pp. 252–256, 2009, doi: 10.1111/j.1469-8986.2008.00769.x.

[34] P. Motyka, M. Grund, N. Forschack, E. Al, A. Villringer, and M. Gaebler, "Interactions between cardiac activity and conscious somatosensory perception," *Psychophysiology*, vol. 56, no. 10, p. e13424, Oct. 2019, doi: 10.1111/psyp.13424.

[35] A. Galvez-Pol, R. McConnell, and J. M. Kilner, "Active sampling in visual search is coupled to the cardiac cycle," *Cognition*, vol. 196, p. 104149, Mar. 2020, doi: 10.1016/j.cognition.2019.104149.

[36] S. Ohl, C. Wohltat, R. Kliegl, O. Pollatos, and R. Engbert, "Microsaccades Are Coupled to Heartbeat," *J. Neurosci.*, vol. 36, no. 4, pp. 1237–1241, Jan. 2016, doi: 10.1523/JNEUROSCI.2211-15.2016.

[37] L. Pramme, M. F. Larra, H. Schächinger, and C. Frings, "Cardiac cycle time effects on selection efficiency in vision," *Psychophysiology*, vol. 53, no. 11, pp. 1702–1711, 2016, doi: 10.1111/psyp.12728.

[38] L. Pramme, M. F. Larra, H. Schächinger, and C. Frings, "Cardiac cycle time effects on mask inhibition," *Biological Psychology*, vol. 100, pp. 115–121, Jul. 2014, doi: 10.1016/j.biopsycho.2014.05.008.

[39] S. Kunzendorf, F. Klotzsche, M. Akbal, A. Villringer, S. Ohl, and M. Gaebler, "Active information sampling varies across the cardiac cycle," *Psychophysiology*, vol. 56, no. 5, p. e13322, 2019, doi: 10.1111/psyp.13322.

[40] A. Galvez-Pol, P. Virdee, J. Villacampa, and J. Kilner, "Active tactile discrimination is coupled with and modulated by the cardiac cycle," *eLife*, vol. 11, p. e78126, Oct. 2022, doi: 10.7554/eLife.78126.

[41] M. F. Larra, J. B. Finke, E. Wascher, and H. Schächinger, "Disentangling sensorimotor and cognitive cardioafferent effects: A cardiac-cycle-time study on spatial stimulus-response compatibility," *Sci Rep*, vol. 10, no. 1, Art. no. 1, Mar. 2020, doi: 10.1038/s41598-020-61068-1.

[42] E. R. Palser, J. Glass, A. Fotopoulou, and J. M. Kilner, "Relationship between cardiac cycle and the timing of actions during action execution and observation," *Cognition*, vol. 217, p. 104907, Diciembre 2021, doi: 10.1016/j.cognition.2021.104907.

[43] C. L. Rae *et al.*, "Response inhibition on the stop signal task improves during cardiac contraction," *Sci Rep*, vol. 8, no. 1, Art. no. 1, Jun. 2018, doi: 10.1038/s41598-018-27513-y.

[44] Q. Ren, A. C. Marshall, J. Kaiser, and S. Schütz-Bosbach, "Response inhibition is disrupted by interoceptive processing at cardiac systole," *Biological Psychology*, vol. 170, p. 108323, Apr. 2022, doi: 10.1016/j.biopsycho.2022.108323.

[45] E. Al, T. Stephani, M. Engelhardt, S. Haegens, A. Villringer, and V. V.

Nikulin, "Cardiac activity impacts cortical motor excitability," *PLOS Biology*, vol. 21, no. 11, p. e3002393, Nov. 2023, doi: 10.1371/journal.pbio.3002393.

[46] E. Al *et al.*, "Heart–brain interactions shape somatosensory perception and evoked potentials," *PNAS*, vol. 117, no. 19, pp. 10575–10584, May 2020, doi: 10.1073/pnas.1915629117.

[47] V. Villani and M. Tsakiris, "Cuspis: A MATLAB Suite for Tasks Investigating Heart-Brain Interactions during fMRI," *Journal of Open Research Software*, vol. 11, no. 1, Art. no. 1, Dec. 2023, doi: 10.5334/jors.476.

[48] R. Schandry, B. Sparrer, and R. Weitkunat, "From the heart to the brain: a study of heartbeat contingent scalp potentials," *Int. J. Neurosci.*, vol. 30, no. 4, pp. 261–275, Nov. 1986.

[49] H.-D. Park and O. Blanke, "Heartbeat-evoked cortical responses: Underlying mechanisms, functional roles, and methodological considerations," *NeuroImage*, vol. 197, pp. 502–511, Aug. 2019, doi: 10.1016/j.neuroimage.2019.04.081.

[50] P. C. Salamone *et al.*, "Dynamic neurocognitive changes in interoception after heart transplant," *Brain Communications*, vol. 2, no. 2, p. fcaa095, Jul. 2020, doi: 10.1093/braincomms/fcaa095.

[51] D. Kumral *et al.*, "Attenuation of the Heartbeat-Evoked Potential in Patients With Atrial Fibrillation," *JACC: Clinical Electrophysiology*, vol. 8, no. 10, pp. 1219–1230, Aug. 2022, doi: 10.1016/j.jacep.2022.06.019.

[52] D. Candia-Rivera *et al.*, "Neural Responses to Heartbeats Detect Residual Signs of Consciousness during Resting State in Postcomatose Patients," *J. Neurosci.*, vol. 41, no. 24, pp. 5251–5262, Jun. 2021, doi: 10.1523/JNEUROSCI.1740-20.2021.

[53] P. C. Salamone *et al.*, "Altered neural signatures of interoception in multiple sclerosis," *Human Brain Mapping*, vol. 39, no. 12, pp. 4743–4754, 2018, doi: 10.1002/hbm.24319.

[54] P. C. Salamone *et al.*, "Interoception Primes Emotional Processing: Multimodal Evidence from Neurodegeneration," *J. Neurosci.*, vol. 41, no. 19, pp. 4276–4292, May 2021, doi: 10.1523/JNEUROSCI.2578-20.2021.

[55] S. Fittipaldi *et al.*, "A multidimensional and multi-feature framework for cardiac interoception," *NeuroImage*, vol. 212, p. 116677, May 2020, doi: 10.1016/j.neuroimage.2020.116677.

[56] M.-P. Coll, H. Hobson, G. Bird, and J. Murphy, "Systematic review and meta-analysis of the relationship between the heartbeat-evoked potential and interoception," *Neuroscience & Biobehavioral Reviews*, vol. 122, pp. 190–200, Mar. 2021, doi: 10.1016/j.neubiorev.2020.12.012.

[57] C. Barà *et al.*, "Local and global measures of information storage for the assessment of heartbeat-evoked cortical responses," *Biomedical Signal Processing and Control*, vol. 86, p. 105315, Sep. 2023, doi: 10.1016/j.bspc.2023.105315.

[58] J. Kim and B. Jeong, "Heartbeat Induces a Cortical Theta-Synchronized Network in the Resting State," *eNeuro*, vol. 6, no. 4, p. ENEURO.0200-19.2019, Jul. 2019, doi: 10.1523/ENEURO.0200-19.2019.

[59] F. Grosselin, X. Navarro-Sune, M. Raux, T. Similowski, and M. Chavez, "CARE-rCortex: A Matlab toolbox for the analysis of CArdio-REspiratory-related activity in the Cortex," *Journal of Neuroscience Methods*, vol. 308, pp. 309–316, Oct. 2018, doi: 10.1016/j.jneumeth.2018.08.011.

[60] W. Lee, E. Kim, J. Park, J. Eo, B. Jeong, and H.-J. Park, "Heartbeat-related spectral perturbation of electroencephalogram reflects dynamic interoceptive attention states in the trial-by-trial classification analysis," *NeuroImage*, p. 120797, Aug. 2024, doi: 10.1016/j.neuroimage.2024.120797.

[61] D. Candia-Rivera and C. Machado, "Multidimensional assessment of heartbeat-evoked responses in disorders of consciousness," *European Journal of Neuroscience*, vol. 58, no. 4, pp. 3098–3110, 2023, doi: 10.1111/ejn.16079.

[62] M. Babo-Rebelo, N. Wolpert, C. Adam, D. Hasboun, and C. Tallon-Baudry, "Is the cardiac monitoring function related to the self in both the default network and right anterior insula?," *Philos. Trans. R. Soc. Lond., B, Biol. Sci.*, vol. 371, no. 1708, Nov. 2016, doi: 10.1098/rstb.2016.0004.

[63] H.-D. Park *et al.*, "Neural Sources and Underlying Mechanisms of Neural Responses to Heartbeats, and their Role in Bodily Self-consciousness: An Intracranial EEG Study," *Cereb Cortex*, vol. 28, no. 7, pp. 2351–2364, Jul. 2018, doi: 10.1093/cercor/bhx136.

[64] U. R. Acharya, K. Paul Joseph, N. Kannathal, C. M. Lim, and J. S. Suri, "Heart rate variability: a review," *Med Biol Eng Comput*, vol. 44, no. 12, pp. 1031–1051, Dec. 2006, doi: 10.1007/s11517-006-0119-0.

[65] R. W. de Boer, J. M. Karemaker, and J. Strackee, "Spectrum of a series of point events, generated by the integral pulse frequency modulation model," *Medical & Biological Engineering & Computing*, vol. 23, no. 2, pp. 138–

142, Mar. 1985, doi: 10.1007/BF02456750.

[66] F. Beissner, K. Meissner, K.-J. Bär, and V. Napadow, "The autonomic brain: an activation likelihood estimation meta-analysis for central processing of autonomic function," *J. Neurosci.*, vol. 33, no. 25, pp. 10503–10511, Jun. 2013, doi: 10.1523/JNEUROSCI.1103-13.2013.

[67] C. Ran, J. C. Boettcher, J. A. Kaye, C. E. Gallori, and S. D. Liberles, "A brainstem map for visceral sensations," *Nature*, pp. 1–7, Aug. 2022, doi: 10.1038/s41586-022-05139-5.

[68] S. Ferraro *et al.*, "The central autonomic system revisited – Convergent evidence for a regulatory role of the insular and midcingulate cortex from neuroimaging meta-analyses," *Neuroscience & Biobehavioral Reviews*, vol. 142, p. 104915, Nov. 2022, doi: 10.1016/j.neubiorev.2022.104915.

[69] L. Mazzola, F. Mauguière, and F. Chouchou, "Central control of cardiac activity as assessed by intra-cerebral recordings and stimulations," *Neurophysiologie Clinique*, vol. 53, no. 2, p. 102849, Apr. 2023, doi: 10.1016/j.neucli.2023.102849.

[70] G. Valenza, L. Passamonti, A. Duggento, N. Toschi, and R. Barbieri, "Uncovering complex central autonomic networks at rest: a functional magnetic resonance imaging study on complex cardiovascular oscillations," *Journal of The Royal Society Interface*, vol. 17, no. 164, p. 20190878, Mar. 2020, doi: 10.1098/rsif.2019.0878.

[71] G. Valenza *et al.*, "The central autonomic network at rest: Uncovering functional MRI correlates of time-varying autonomic outflow," *NeuroImage*, vol. 197, pp. 383–390, Aug. 2019, doi: 10.1016/j.neuroimage.2019.04.075.

[72] A. Duggento *et al.*, "Globally conditioned Granger causality in brain–brain and brain–heart interactions: a combined heart rate variability/ultra-high-field (7 T) functional magnetic resonance imaging study," *Philosophical Transactions of the Royal Society A: Mathematical, Physical and Engineering Sciences*, vol. 374, no. 2067, p. 20150185, May 2016, doi: 10.1098/rsta.2015.0185.

[73] A. Bashan, R. P. Bartsch, J. W. Kantelhardt, S. Havlin, and P. C. Ivanov, "Network physiology reveals relations between network topology and physiological function," *Nature Communications*, vol. 3, no. 1, Art. no. 1, Feb. 2012, doi: 10.1038/ncomms1705.

[74] R. Pernice *et al.*, "Multivariate Correlation Measures Reveal Structure and Strength of Brain–Body Physiological Networks at Rest and During Mental Stress," *Frontiers in Neuroscience*, vol. 14, p. 1427, 2021, doi: 10.3389/fnins.2020.602584.

[75] M. Ako *et al.*, "Correlation between electroencephalography and heart rate variability during sleep," *Psychiatry Clin Neurosci*, vol. 57, no. 1, pp. 59–65, Feb. 2003, doi: 10.1046/j.1440-1819.2003.01080.x.

[76] L. Faes *et al.*, "Information decomposition in the frequency domain: a new framework to study cardiovascular and cardiorespiratory oscillations," *Philosophical Transactions of the Royal Society A: Mathematical, Physical and Engineering Sciences*, vol. 379, no. 2212, p. 20200250, Oct. 2021, doi: 10.1098/rsta.2020.0250.

[77] R. Pernice *et al.*, "Spectral decomposition of cerebrovascular and cardiovascular interactions in patients prone to postural syncope and healthy controls," *Autonomic Neuroscience*, vol. 242, p. 103021, Nov. 2022, doi: 10.1016/j.autneu.2022.103021.

[78] D. N. Reshef *et al.*, "Detecting Novel Associations in Large Datasets," *Science*, vol. 334, no. 6062, pp. 1518–1524, Dec. 2011, doi: 10.1126/science.1205438.

[79] C. J. Stam and B. W. van Dijk, "Synchronization likelihood: an unbiased measure of generalized synchronization in multivariate data sets," *Physica D: Nonlinear Phenomena*, vol. 163, no. 3, pp. 236–251, Mar. 2002, doi: 10.1016/S0167-2789(01)00386-4.

[80] M. Dumont, F. Jurysta, J.-P. Lanquart, P.-F. Migeotte, P. van de Borne, and P. Linkowski, "Interdependency between heart rate variability and sleep EEG: linear/non-linear?," *Clinical Neurophysiology*, vol. 115, no. 9, pp. 2031–2040, Sep. 2004, doi: 10.1016/j.clinph.2004.04.007.

[81] V. Catrambone, D. Candia-Rivera, and G. Valenza, "Intracortical brain-heart interplay: An EEG model source study of sympathovagal changes," *Human Brain Mapping*, vol. 45, no. 6, p. e26677, 2024, doi: 10.1002/hbm.26677.

[82] S. Schulz, M. Bolz, K.-J. Bär, and A. Voss, "Central- and autonomic nervous system coupling in schizophrenia," *Philosophical Transactions of the Royal Society A: Mathematical, Physical and Engineering Sciences*, vol. 374, no. 2067, p. 20150178, May 2016, doi: 10.1098/rsta.2015.0178.

[83] F. Jurysta *et al.*, "A study of the dynamic interactions between sleep EEG and heart rate variability in healthy young men," *Clin Neurophysiol*, vol. 114, no. 11, pp. 2146–2155, Nov. 2003, doi: 10.1016/s1388-2457(03)00215-3.

[84] D. Li *et al.*, "Asphyxia-activated corticocardiac signaling accelerates onset

of cardiac arrest," *Proc Natl Acad Sci U S A*, vol. 112, no. 16, pp. E2073-2082, Apr. 2015, doi: 10.1073/pnas.1423936112.

[85] M. M. Admiraal, E. J. Gilmore, M. J. A. M. Van Putten, H. P. Zaveri, L. J. Hirsch, and N. Gaspard, "Disruption of Brain–Heart Coupling in Sepsis," *Journal of Clinical Neurophysiology*, vol. 34, no. 5, pp. 413–420, Sep. 2017, doi: 10.1097/WNP.0000000000000381.

[86] S. Schulz, N. Tupaika, S. Berger, J. Haueisen, K.-J. Bär, and A. Voss, "Cardiovascular coupling analysis with high-resolution joint symbolic dynamics in patients suffering from acute schizophrenia," *Physiol Meas*, vol. 34, no. 8, pp. 883–901, Aug. 2013, doi: 10.1088/0967-3334/34/8/883.

[87] M. Wibral, R. Vicente, and J. T. Lizier, Eds., *Directed Information Measures in Neuroscience*. in Understanding Complex Systems. Berlin, Heidelberg: Springer, 2014. doi: 10.1007/978-3-642-54474-3.

[88] C. Barà *et al.*, "Comparison of discretization strategies for the model-free information-theoretic assessment of short-term physiological interactions," *Chaos: An Interdisciplinary Journal of Nonlinear Science*, vol. 33, no. 3, p. 033127, Mar. 2023, doi: 10.1063/5.0140641.

[89] Y. Antonacci, C. Barà, A. Zaccaro, F. Ferri, R. Pernice, and L. Faes, "Time-varying information measures: an adaptive estimation of information storage with application to brain-heart interactions," *Front Netw Physiol*, vol. 3, p. 1242505, Oct. 2023, doi: 10.3389/fnetp.2023.1242505.

[90] M. Wibral, V. Priesemann, J. W. Kay, J. T. Lizier, and W. A. Phillips, "Partial information decomposition as a unified approach to the specification of neural goal functions," *Brain and Cognition*, vol. 112, pp. 25–38, Mar. 2017, doi: 10.1016/j.bandc.2015.09.004.

[91] G. Sugihara *et al.*, "Detecting Causality in Complex Ecosystems," *Science*, vol. 338, no. 6106, pp. 496–500, Oct. 2012, doi: 10.1126/science.1227079.

[92] J. Arnhold, P. Grassberger, K. Lehnertz, and C. E. Elger, "A robust method for detecting interdependences: application to intracranially recorded EEG," *Physica D: Nonlinear Phenomena*, vol. 134, no. 4, pp. 419–430, Dec. 1999, doi: 10.1016/S0167-2789(99)00140-2.

[93] M. Le Van Quyen, J. Martinerie, C. Adam, and F. J. Varela, "Nonlinear analyses of interictal EEG map the brain interdependences in human focal epilepsy," *Physica D: Nonlinear Phenomena*, vol. 127, no. 3, pp. 250–266, Mar. 1999, doi: 10.1016/S0167-2789(98)00258-9.

[94] M. Breakspear and J. R. Terry, "Detection and description of non-linear interdependence in normal multichannel human EEG data," *Clinical Neurophysiology*, vol. 113, no. 5, pp. 735–753, May 2002, doi: 10.1016/S1388-2457(02)00051-2.

[95] J. R. Terry and M. Breakspear, "An improved algorithm for the detection of dynamical interdependence in bivariate time-series," *Biol Cybern*, vol. 88, no. 2, pp. 129–136, Feb. 2003, doi: 10.1007/s00422-002-0368-4.

[96] E. Pereda, R. Rial, A. Gamundi, and J. González, "Assessment of changing interdependencies between human electroencephalograms using nonlinear methods," *Physica D: Nonlinear Phenomena*, vol. 148, no. 1, pp. 147–158, Jan. 2001, doi: 10.1016/S0167-2789(00)00190-1.

[97] D. Chicharro, "On the spectral formulation of Granger causality," *Biol Cybern*, vol. 105, no. 5, pp. 331–347, Dec. 2011, doi: 10.1007/s00422-011-0469-z.

[98] A. M. Bastos and J.-M. Schoffelen, "A Tutorial Review of Functional Connectivity Analysis Methods and Their Interpretational Pitfalls," *Frontiers in Systems Neuroscience*, vol. 9, 2016, Accessed: Nov. 09, 2022. [Online]. Available: https://www.frontiersin.org/articles/10.3389/fnsys.2015.00175

[99] A. Porta and L. Faes, "Wiener–Granger Causality in Network Physiology With Applications to Cardiovascular Control and Neuroscience," *Proceedings of the IEEE*, vol. 104, no. 2, pp. 282–309, Feb. 2016, doi: 10.1109/JPROC.2015.2476824.

[100] S. J. Schiff, P. So, T. Chang, R. E. Burke, and T. Sauer, "Detecting dynamical interdependence and generalized synchrony through mutual prediction in a neural ensemble," *Phys. Rev. E*, vol. 54, no. 6, pp. 6708–6724, Dec. 1996, doi: 10.1103/PhysRevE.54.6708.

[101] A. Porta *et al.*, "On the validity of the state space correspondence strategy based on k-nearest neighbor cross-predictability in assessing directionality in stochastic systems: Application to cardiorespiratory coupling estimation," *Chaos*, vol. 34, no. 5, p. 053115, May 2024, doi: 10.1063/5.0192645.

[102] A. Müller, J. F. Kraemer, T. Penzel, H. Bonnemeier, J. Kurths, and N. Wessel, "Causality in physiological signals," *Physiol. Meas.*, vol. 37, no. 5, p. R46, Apr. 2016, doi: 10.1088/0967-3334/37/5/R46.

[103] C. W. J. Granger, "Investigating Causal Relations by Econometric Models and Cross-spectral Methods," *Econometrica*, vol. 37, no. 3, pp. 424–438, 1969, doi: 10.2307/1912791.

[104] L. Faes, G. Nollo, and K. H. Chon, "Assessment of Granger causality by nonlinear model identification: application to short-term cardiovascular

variability," *Ann Biomed Eng*, vol. 36, no. 3, pp. 381–395, Mar. 2008, doi: 10.1007/s10439-008-9441-z.

[105] D. Marinazzo, M. Pellicoro, and S. Stramaglia, "Kernel Method for Nonlinear Granger Causality," *Phys. Rev. Lett.*, vol. 100, no. 14, p. 144103, Apr. 2008, doi: 10.1103/PhysRevLett.100.144103.

[106] K. Friston, R. Moran, and A. K. Seth, "Analysing connectivity with Granger causality and dynamic causal modelling," *Current Opinion in Neurobiology*, vol. 23, no. 2, pp. 172–178, Apr. 2013, doi: 10.1016/j.conb.2012.11.010.

[107] S. L. Bressler and A. K. Seth, "Wiener–Granger Causality: A well established methodology," *NeuroImage*, vol. 58, no. 2, pp. 323–329, Sep. 2011, doi: 10.1016/j.neuroimage.2010.02.059.

[108] S. Schulz *et al.*, "Cardiovascular and cardiorespiratory coupling analyses: a review," *Philosophical Transactions of the Royal Society A: Mathematical, Physical and Engineering Sciences*, vol. 371, no. 1997, p. 20120191, Aug. 2013, doi: 10.1098/rsta.2012.0191.

[109] M. Javorka, B. Czippelova, Z. Turianikova, Z. Lazarova, I. Tonhajzerova, and L. Faes, "Causal analysis of short-term cardiovascular variability: state-dependent contribution of feedback and feedforward mechanisms," *Med Biol Eng Comput*, vol. 55, no. 2, pp. 179–190, Feb. 2017, doi: 10.1007/s11517-016-1492-y.

[110] A. Duggento, M. Guerrisi, and N. Toschi, "Echo state network models for nonlinear Granger causality," *Philosophical Transactions of the Royal Society A: Mathematical, Physical and Engineering Sciences*, vol. 379, no. 2212, p. 20200256, Dec. 2021, doi: 10.1098/rsta.2020.0256.

[111] A. Greco, L. Faes, V. Catrambone, R. Barbieri, E. P. Scilingo, and G. Valenza, "Lateralization of directional brain-heart information transfer during visual emotional elicitation," *American Journal of Physiology-Regulatory, Integrative and Comparative Physiology*, vol. 317, no. 1, pp. R25–R38, May 2019, doi: 10.1152/ajpregu.00151.2018.

[112] A. D. Orjuela-Cañón, A. Cerquera, J. A. Freund, G. Juliá-Serdá, and A. G. Ravelo-García, "Sleep apnea: Tracking effects of a first session of CPAP therapy by means of Granger causality," *Computer Methods and Programs in Biomedicine*, vol. 187, p. 105235, Apr. 2020, doi: 10.1016/j.cmpb.2019.105235.

[113] L. Faes, D. Marinazzo, S. Stramaglia, F. Jurysta, A. Porta, and N. Giandomenico, "Predictability decomposition detects the impairment of brain-heart dynamical networks during sleep disorders and their recovery with treatment," *Philos Trans A Math Phys Eng Sci*, vol. 374, no. 2067, p. 20150177, May 2016, doi: 10.1098/rsta.2015.0177.

[114] R. Pernice, L. Faes, M. Feucht, F. Benninger, S. Mangione, and K. Schiecke, "Pairwise and higher-order measures of brain-heart interactions in children with temporal lobe epilepsy," *J. Neural Eng.*, vol. 19, no. 4, p. 045002, Jul. 2022, doi: 10.1088/1741-2552/ac7fba.

[115] T. Schreiber, "Measuring Information Transfer," *Phys. Rev. Lett.*, vol. 85, no. 2, pp. 461–464, Jul. 2000, doi: 10.1103/PhysRevLett.85.461.

[116] L. Faes, G. Nollo, and A. Porta, "Information-based detection of nonlinear Granger causality in multivariate processes via a nonuniform embedding technique," *Phys. Rev. E*, vol. 83, no. 5, p. 051112, May 2011, doi: 10.1103/PhysRevE.83.051112.

[117] R. Vicente, M. Wibral, M. Lindner, and G. Pipa, "Transfer entropy—a model-free measure of effective connectivity for the neurosciences," *J Comput Neurosci*, vol. 30, no. 1, pp. 45–67, Feb. 2011, doi: 10.1007/s10827-010-0262-3.

[118] X. Yu, C. Zhang, L. Su, J. Zhang, and N. Rao, "Estimation of the cortico-cortical and brain-heart functional coupling with directed transfer function and corrected conditional entropy," *Biomedical Signal Processing and Control*, vol. 43, pp. 110–116, May 2018, doi: 10.1016/j.bspc.2018.01.018.

[119] M. Kumar, D. Singh, and K. K. Deepak, "Identifying heart-brain interactions during internally and externally operative attention using conditional entropy," *Biomedical Signal Processing and Control*, vol. 57, p. 101826, Mar. 2020, doi: 10.1016/j.bspc.2019.101826.

[120] K. Schiecke, L. Leistritz, and L. Iasemidis, "Brain-Heart Interactions in Long-Term Recordings of Epileptic Patients: Proof-of-Principle Application of Segmented Convergent Cross Mapping," in *2020 11th Conference of the European Study Group on Cardiovascular Oscillations (ESGCO)*, Jul. 2020, pp. 1–2. doi: 10.1109/ESGCO49734.2020.9158051.

[121] K. Schiecke, F. Benninger, and M. Feucht, "Analysis of Brain-Heart Couplings in Epilepsy: Dealing With the Highly Complex Structure of Resulting Interaction Pattern," in *2020 28th European Signal Processing Conference (EUSIPCO)*, Jan. 2021, pp. 935–939. doi: 10.23919/Eusipco47968.2020.9287620.

[122] L. Frassineti, C. Manfredi, D. Ermini, R. Fabbri, B. Olmi, and A. Lanatà, "Analysis of Brain-Heart Interactions in newborns with and without seizures using the Convergent Cross Mapping approach," in *2022 44th*

*Annual International Conference of the IEEE Engineering in Medicine & Biology Society (EMBC)*, Jul. 2022, pp. 36–39. doi: 10.1109/EMBC48229.2022.9871141.

[123] K. Schiecke *et al.*, "Nonlinear Directed Interactions Between HRV and EEG Activity in Children With TLE," *IEEE Transactions on Biomedical Engineering*, vol. 63, no. 12, pp. 2497–2504, Dec. 2016, doi: 10.1109/TBME.2016.2579021.

[124] D. Candia-Rivera, "Modeling brain-heart interactions from Poincaré plot-derived measures of sympathetic-vagal activity," *MethodsX*, vol. 10, p. 102116, Mar. 2023, doi: 10.1016/j.mex.2023.102116.

[125] R. Bailón, G. Laouini, C. Grao, M. Orini, P. Laguna, and O. Meste, "The integral pulse frequency modulation model with time-varying threshold: application to heart rate variability analysis during exercise stress testing," *IEEE Trans Biomed Eng*, vol. 58, no. 3, pp. 642–652, Mar. 2011, doi: 10.1109/TBME.2010.2095011.

[126] J. D. Scheff, P. D. Mavroudis, S. E. Calvano, S. F. Lowry, and I. P. Androulakis, "Modeling autonomic regulation of cardiac function and heart rate variability in human endotoxemia," *Physiol Genomics*, vol. 43, no. 16, pp. 951–964, Aug. 2011, doi: 10.1152/physiolgenomics.00040.2011.

[127] D. C. McLernon, N. J. Dabanloo, A. Ayatollahi, V. J. Majd, and H. Zhang, "A new nonlinear model for generating RR tachograms," in *Computers in Cardiology, 2004*, Sep. 2004, pp. 481–484. doi: 10.1109/CIC.2004.1442979.

[128] H.-W. Chiu, T.-H. Wang, L.-C. Huang, H.-W. Tso, and T. Kao, "The influence of mean heart rate on measures of heart rate variability as markers of autonomic function: a model study," *Med Eng Phys*, vol. 25, no. 6, pp. 475–481, Jul. 2003, doi: 10.1016/s1350-4533(03)00019-5.

[129] B. W. Hyndman and R. K. Mohn, "A model of the cardiac pacemaker and its use in decoding the information content of cardiac intervals," *Automedica*, vol. 1, pp. 239–252, 1975.

[130] D. Candia-Rivera, V. Catrambone, R. Barbieri, and G. Valenza, "Integral pulse frequency modulation model driven by sympathovagal dynamics: Synthetic vs. real heart rate variability," *Biomedical Signal Processing and Control*, vol. 68, p. 102736, Jul. 2021, doi: 10.1016/j.bspc.2021.102736.

[131] H. Al-Nashash, Y. Al-Assaf, J. Paul, and N. Thakor, "EEG signal modeling using adaptive Markov process amplitude," *IEEE Transactions on Biomedical Engineering*, vol. 51, no. 5, pp. 744–751, May 2004, doi: 10.1109/TBME.2004.826602.

[132] D. Candia-Rivera, V. Catrambone, J. F. Thayer, C. Gentili, and G. Valenza, "Cardiac sympathetic-vagal activity initiates a functional brain–body response to emotional arousal," *Proceedings of the National Academy of Sciences*, vol. 119, no. 21, p. e2119599119, May 2022, doi: 10.1073/pnas.2119599119.

[133] D. Candia-Rivera, K. Norouzi, T. Z. Ramsøy, and G. Valenza, "Dynamic fluctuations in ascending heart-to-brain communication under mental stress," *American Journal of Physiology-Regulatory, Integrative and Comparative Physiology*, vol. 324, no. 4, pp. R513–R525, Apr. 2023, doi: 10.1152/ajpregu.00251.2022.

[134] V. Catrambone, S. Messerotti Benvenuti, C. Gentili, and G. Valenza, "Intensification of functional neural control on heartbeat dynamics in subclinical depression," *Translational Psychiatry*, vol. 11, no. 1, Art. no. 1, Apr. 2021, doi: 10.1038/s41398-021-01336-4.

[135] B. Hermann *et al.*, "Aberrant brain–heart coupling is associated with the severity of post cardiac arrest brain injury," *Annals of Clinical and Translational Neurology*, vol. 11, no. 4, pp. 866–882, 2024, doi: 10.1002/acn3.52000.

[136] F. Battiston *et al.*, "Networks beyond pairwise interactions: Structure and dynamics," *Physics Reports*, vol. 874, pp. 1–92, Aug. 2020, doi: 10.1016/j.physrep.2020.05.004.

[137] A. I. Luppi, F. E. Rosas, P. A. M. Mediano, D. K. Menon, and E. A. Stamatakis, "Information decomposition and the informational architecture of the brain," *Trends in Cognitive Sciences*, Jan. 2024, doi: 10.1016/j.tics.2023.11.005.

[138] W. McGill, "Multivariate information transmission," *Transactions on the IRE Professional Group on Information Theory*, vol. 4, no. 4, pp. 93–111, Sep. 1954, doi: 10.1109/TIT.1954.1057469.

[139] F. E. Rosas, P. A. M. Mediano, M. Gastpar, and H. J. Jensen, "Quantifying high-order interdependencies via multivariate extensions of the mutual information," *Phys. Rev. E*, vol. 100, no. 3, p. 032305, Sep. 2019, doi: 10.1103/PhysRevE.100.032305.

[140] J. T. Lizier, N. Bertschinger, J. Jost, and M. Wibral, "Information Decomposition of Target Effects from Multi-Source Interactions: Perspectives on Previous, Current and Future Work," *Entropy*, vol. 20, no. 4, Art. no. 4, Apr. 2018, doi: 10.3390/e20040307.

[141] M. Javorka *et al.*, "Towards understanding the complexity of cardiovascular oscillations: Insights from information theory," *Comput Biol Med*, vol. 98, pp. 48–57, Jul. 2018, doi: 10.1016/j.compbiomed.2018.05.007.

[142] T. F. Varley, M. Pope, J. Faskowitz, and O. Sporns, "Multivariate information theory uncovers synergistic subsystems of the human cerebral cortex," *Commun Biol*, vol. 6, no. 1, pp. 1–12, Apr. 2023, doi: 10.1038/s42003-023-04843-w.

[143] L. Faes *et al.*, "A New Framework for the Time- and Frequency-Domain Assessment of High-Order Interactions in Networks of Random Processes," *IEEE Transactions on Signal Processing*, vol. 70, pp. 5766–5777, 2022, doi: 10.1109/TSP.2022.3221892.

[144] L. Sparacino, Y. Antonacci, G. Mijatovic, and L. Faes, "Measuring hierarchically-organized interactions in dynamic networks through spectral entropy rates: theory, estimation, and illustrative application to physiological networks," Jan. 20, 2024, *arXiv*: arXiv:2401.11327. doi: 10.48550/arXiv.2401.11327.

[145] T. Scagliarini *et al.*, "Gradients of O-information: Low-order descriptors of high-order dependencies," *Phys. Rev. Res.*, vol. 5, no. 1, p. 013025, Jan. 2023, doi: 10.1103/PhysRevResearch.5.013025.

[146] G. Mijatovic *et al.*, "Assessing High-Order Links in Cardiovascular and Respiratory Networks via Static and Dynamic Information Measures," Jan. 10, 2024, *arXiv*: arXiv:2401.05556. doi: 10.48550/arXiv.2401.05556.

[147] S. L. Bressler and V. Menon, "Large-scale brain networks in cognition: emerging methods and principles," *Trends in Cognitive Sciences*, vol. 14, no. 6, pp. 277–290, Jun. 2010, doi: 10.1016/j.tics.2010.04.004.

[148] H.-J. Park and K. Friston, "Structural and Functional Brain Networks: From Connections to Cognition," *Science*, vol. 342, no. 6158, p. 1238411, Nov. 2013, doi: 10.1126/science.1238411.

[149] C. Chang, C. D. Metzger, G. H. Glover, J. H. Duyn, H.-J. Heinze, and M. Walter, "Association between heart rate variability and fluctuations in resting-state functional connectivity," *NeuroImage*, vol. 68, pp. 93–104, Mar. 2013, doi: 10.1016/j.neuroimage.2012.11.038.

[150] D. Kumral *et al.*, "The age-dependent relationship between resting heart rate variability and functional brain connectivity," *NeuroImage*, vol. 185, pp. 521–533, Jan. 2019, doi: 10.1016/j.neuroimage.2018.10.027.

[151] S. D. X. Kong *et al.*, "Heart rate variability during slow wave sleep is linked to functional connectivity in the central autonomic network," *Brain Communications*, vol. 5, no. 3, p. fcad129, Jun. 2023, doi: 10.1093/braincomms/fcad129.

[152] D. Candia-Rivera, M. Vidailhet, M. Chavez, and F. De Vico Fallani, "A framework for quantifying the coupling between brain connectivity and heartbeat dynamics: Insights into the disrupted network physiology in Parkinson's disease," *Human Brain Mapping*, vol. 45, no. 5, p. e26668, 2024, doi: 10.1002/hbm.26668.

[153] D. Candia-Rivera, M. Chavez, and F. de Vico Fallani, "Measures of the coupling between fluctuating brain network organization and heartbeat dynamics," *Network Neuroscience*, vol. 8, no. 2, pp. 557–575, Mar. 2024, doi: 10.1162/netn_a_00369.

[154] M. Rubinov and O. Sporns, "Complex network measures of brain connectivity: Uses and interpretations," *NeuroImage*, vol. 52, no. 3, pp. 1059–1069, Sep. 2010, doi: 10.1016/j.neuroimage.2009.10.003.

[155] C. M. Signorelli, J. D. Boils, E. Tagliazucchi, B. Jarraya, and G. Deco, "From Brain-Body Function to Conscious Interactions," *Neuroscience & Biobehavioral Reviews*, vol. 141, p. 104833, Aug. 2022, doi: 10.1016/j.neubiorev.2022.104833.

[156] C. Presigny and F. De Vico Fallani, "Colloquium: Multiscale modeling of brain network organization," *Rev. Mod. Phys.*, vol. 94, no. 3, p. 031002, Aug. 2022, doi: 10.1103/RevModPhys.94.031002.

[157] Task Force of the European Society of Cardiology the North American Society of Pacing, "Heart Rate Variability: Standards of Measurement, Physiological Interpretation, and Clinical Use," *Circulation*, vol. 93, no. 5, pp. 1043–1065, Mar. 1996, doi: 10.1161/01.CIR.93.5.1043.

[158] J.-M. Schoffelen and J. Gross, "Source connectivity analysis with MEG and EEG," *Hum Brain Mapp*, vol. 30, no. 6, pp. 1857–1865, Jun. 2009, doi: 10.1002/hbm.20745.

[159] T. Demiralp, J. Yordanova, V. Kolev, A. Ademoglu, M. Devrim, and V. J. Samar, "Time–Frequency Analysis of Single-Sweep Event-Related Potentials by Means of Fast Wavelet Transform," *Brain and Language*, vol. 66, no. 1, pp. 129–145, Jan. 1999, doi: 10.1006/brln.1998.2028.

[160] M. E. Tağluk, E. D. Çakmak, and S. Karakaş, "Analysis of the time-varying energy of brain responses to an oddball paradigm using short-term smoothed Wigner–Ville distribution," *Journal of Neuroscience Methods*, vol. 143, no. 2, pp. 197–208, Apr. 2005, doi: 10.1016/j.jneumeth.2004.10.007.

[161] S. S. Khalsa *et al.*, "Interoception and Mental Health: A Roadmap," *Biological Psychiatry: Cognitive Neuroscience and Neuroimaging*, vol. 3, no. 6, pp. 501–513, Jun. 2018, doi: 10.1016/j.bpsc.2017.12.004.

[162] S. N. Garfinkel, C. Tiley, S. O'Keeffe, N. A. Harrison, A. K. Seth, and H. D. Critchley, "Discrepancies between dimensions of interoception in autism: Implications for emotion and anxiety," *Biological Psychology*, vol. 114, pp. 117–126, Feb. 2016, doi: 10.1016/j.biopsycho.2015.12.003.

[163] E. F. Pace-Schott *et al.*, "Physiological feelings," *Neuroscience & Biobehavioral Reviews*, vol. 103, pp. 267–304, Aug. 2019, doi: 10.1016/j.neubiorev.2019.05.002.

[164] N. R. Winter *et al.*, "A Systematic Evaluation of Machine Learning–Based Biomarkers for Major Depressive Disorder," *JAMA Psychiatry*, Jan. 2024, doi: 10.1001/jamapsychiatry.2023.5083.

[165] B. W. J. H. Penninx, "Depression and cardiovascular disease: Epidemiological evidence on their linking mechanisms," *Neuroscience & Biobehavioral Reviews*, vol. 74, pp. 277–286, Mar. 2017, doi: 10.1016/j.neubiorev.2016.07.003.

[166] W. Whang *et al.*, "Depression and Risk of Sudden Cardiac Death and Coronary Heart Disease in Women," *Journal of the American College of Cardiology*, vol. 53, no. 11, pp. 950–958, Mar. 2009, doi: 10.1016/j.jacc.2008.10.060.

[167] E. Patron *et al.*, "Association between depression and heart rate variability in patients after cardiac surgery: A pilot study," *Journal of Psychosomatic Research*, vol. 73, no. 1, pp. 42–46, Jul. 2012, doi: 10.1016/j.jpsychores.2012.04.013.

[168] L. Quadt, H. D. Critchley, and S. N. Garfinkel, "The neurobiology of interoception in health and disease," *Annals of the New York Academy of Sciences*, vol. 1428, no. 1, pp. 112–128, 2018, doi: 10.1111/nyas.13915.

[169] R. G. Garcia *et al.*, "Impact of sex and depressed mood on the central regulation of cardiac autonomic function," *Neuropsychopharmacol.*, vol. 45, no. 8, Art. no. 8, Jul. 2020, doi: 10.1038/s41386-020-0651-x.

[170] V. Catrambone, L. Zallocco, E. Ramoretti, M. R. Mazzoni, L. Sebastiani, and G. Valenza, "Integrative Neuro-Cardiovascular Dynamics in Response to Test Anxiety: a Brain-Heart Axis study," *Physiology & Behavior*, p. 114460, Jan. 2024, doi: 10.1016/j.physbeh.2024.114460.

[171] T. A. Iseger, N. E. R. van Bueren, J. L. Kenemans, R. Gevirtz, and M. Arns, "A frontal-vagal network theory for Major Depressive Disorder: Implications for optimizing neuromodulation techniques," *Brain Stimulation: Basic, Translational, and Clinical Research in Neuromodulation*, vol. 13, no. 1, pp. 1–9, Jan. 2020, doi: 10.1016/j.brs.2019.10.006.

[172] T. A. Iseger, F. Padberg, J. L. Kenemans, H. van Dijk, and M. Arns, "Neuro-Cardiac-Guided TMS (NCG TMS): A replication and extension study," *Biological Psychology*, vol. 162, p. 108097, May 2021, doi: 10.1016/j.biopsycho.2021.108097.

[173] S. Jain, "Multi-organ autonomic dysfunction in Parkinson disease," *Parkinsonism & Related Disorders*, vol. 17, no. 2, pp. 77–83, Feb. 2011, doi: 10.1016/j.parkreldis.2010.08.022.

[174] Y. Sharabi, G. D. Vatine, and A. Ashkenazi, "Parkinson's disease outside the brain: targeting the autonomic nervous system," *The Lancet Neurology*, vol. 20, no. 10, pp. 868–876, Oct. 2021, doi: 10.1016/S1474-4422(21)00219-2.

[175] L. Ricciardi *et al.*, "Know thyself: Exploring interoceptive sensitivity in Parkinson's disease," *Journal of the Neurological Sciences*, vol. 364, pp. 110–115, May 2016, doi: 10.1016/j.jns.2016.03.019.

[176] G. Santangelo *et al.*, "Interoceptive processing deficit: A behavioral marker for subtyping Parkinson's disease," *Parkinsonism & Related Disorders*, vol. 53, pp. 64–69, Aug. 2018, doi: 10.1016/j.parkreldis.2018.05.001.

[177] J. L. Hazelton *et al.*, "Thinking versus feeling: How interoception and cognition influence emotion recognition in behavioural-variant frontotemporal dementia, Alzheimer's disease, and Parkinson's disease,"

[178] M. Iniguez *et al.*, "Heart-brain synchronization breakdown in Parkinson's disease," *NPJ Parkinsons Dis*, vol. 8, no. 1, p. 64, May 2022, doi: 10.1038/s41531-022-00323-w.

[179] L. Lin, Y. Cheng, P. Huang, J. Zhang, J. Zheng, and X. Pan, "Synchronous monitoring of brain-heart electrophysiology using heart rate variability coupled with rapid quantitative electroencephalography in orthostatic hypotension patients with α-synucleinopathies: Rapid prediction of orthostatic hypotension and preliminary exploration of brain stimulation therapy," *CNS Neuroscience & Therapeutics*, vol. 30, no. 2, p. e14571, 2024, doi: 10.1111/cns.14571.

[180] J. Agrimi *et al.*, "Cardiac AC8 Over-Expression Increases Locomotion by Altering Heart-Brain Communication," *JACC: Clinical Electrophysiology*, Sep. 2023, doi: 10.1016/j.jacep.2023.07.023.

[181] J. J. Daly and J. R. Wolpaw, "Brain–computer interfaces in neurological rehabilitation," *The Lancet Neurology*, vol. 7, no. 11, pp. 1032–1043, Nov. 2008, doi: 10.1016/S1474-4422(08)70223-0.

[182] G. Naros, I. Naros, F. Grimm, U. Ziemann, and A. Gharabaghi, "Reinforcement learning of self-regulated sensorimotor β-oscillations improves motor performance," *NeuroImage*, vol. 134, pp. 142–152, Jul. 2016, doi: 10.1016/j.neuroimage.2016.03.016.

[183] C. Sannelli, C. Vidaurre, K.-R. Müller, and B. Blankertz, "A large scale screening study with a SMR-based BCI: Categorization of BCI users and differences in their SMR activity," *PLOS ONE*, vol. 14, no. 1, p. e0207351, Jan. 2019, doi: 10.1371/journal.pone.0207351.

[184] A. Ghouse, D. Candia-Rivera, and G. Valenza, "Nonlinear neural patterns are revealed in high frequency functional near infrared spectroscopy analysis," *Brain Research Bulletin*, vol. 203, p. 110759, Oct. 2023, doi: 10.1016/j.brainresbull.2023.110759.

[185] V. Catrambone, G. Averta, M. Bianchi, and G. Valenza, "Toward brain–heart computer interfaces: a study on the classification of upper limb movements using multisystem directional estimates," *J. Neural Eng.*, vol. 18, no. 4, p. 046002, Mar. 2021, doi: 10.1088/1741-2552/abe7b9.

[186] R. Mane, T. Chouhan, and C. Guan, "BCI for stroke rehabilitation: motor and beyond," *J. Neural Eng.*, vol. 17, no. 4, p. 041001, Aug. 2020, doi: 10.1088/1741-2552/aba162.

[187] J. T. Giacino, K. Kalmar, and J. Whyte, "The JFK Coma Recovery Scale-Revised: measurement characteristics and diagnostic utility," *Arch Phys Med Rehabil*, vol. 85, no. 12, pp. 2020–2029, Dicembre 2004.

[188] D. Kondziella *et al.*, "European Academy of Neurology guideline on the diagnosis of coma and other disorders of consciousness," *European Journal of Neurology*, vol. 27, no. 5, pp. 741–756, 2020, doi: 10.1111/ene.14151.

[189] B. Hermann *et al.*, "Importance, limits and caveats of the use of 'disorders of consciousness' to theorize consciousness," *Neuroscience of Consciousness*, vol. 2021, no. 2, p. niab048, Dec. 2021, doi: 10.1093/nc/niab048.

[190] L. R. D. Sanz, A. Thibaut, B. L. Edlow, S. Laureys, and O. Gosseries, "Update on neuroimaging in disorders of consciousness," *Curr Opin Neurol*, vol. 34, no. 4, pp. 488–496, Aug. 2021, doi: 10.1097/WCO.0000000000000951.

[191] F. Raimondo *et al.*, "Brain–heart interactions reveal consciousness in noncommunicating patients," *Annals of Neurology*, vol. 82, no. 4, pp. 578–591, 2017, doi: 10.1002/ana.25045.

[192] D. Candia-Rivera, F. Raimondo, P. Pérez, L. Naccache, C. Tallon-Baudry, and J. D. Sitt, "Conscious processing of global and local auditory irregularities causes differentiated heartbeat-evoked responses," *eLife*, vol. 12, p. e75352, Oct. 2023, doi: 10.7554/eLife.75352.

[193] P. Pérez *et al.*, "Conscious processing of narrative stimuli synchronizes heart rate between individuals," *Cell Rep*, vol. 36, no. 11, p. 109692, 2021, doi: 10.1016/j.celrep.2021.109692.

*Cortex*, vol. 163, pp. 66–79, Mar. 2023, doi: 10.1016/j.cortex.2023.02.009.